\documentclass[twocolumn,tighten,times]{aastex631} % linenumbers,
\usepackage{float}
\usepackage{graphicx}
\usepackage{amssymb}
\usepackage{amsbsy}
\usepackage{amsmath}
\usepackage{amsfonts}
\usepackage{mathrsfs}
\usepackage{color}
\usepackage{multirow}
\usepackage{bm}

\DeclareMathAlphabet{\mathbi}{OT1}{ptm}{bx}{it}
\SetMathAlphabet\mathbi{bold}{OT1}{ptm}{bx}{it}

\shorttitle{SMBH and BLR in NGC~5548}
\shortauthors{Lu et al.}

\begin{document}

\title{\bf\large Supermassive Black Hole and Broad-line Region in NGC~5548: Results from Five-season Reverberation Mapping}

\author[0000-0002-2310-0982]{Kai-Xing Lu}
\affiliation{Yunnan Observatories, Chinese Academy of Sciences, Kunming 650011, People's Republic of China} 
\affiliation{Key Laboratory for the Structure and Evolution of Celestial Objects, Chinese Academy of Sciences, Kunming 650011, People's Republic of China} 
\author{Jin-Ming Bai}
\affiliation{Yunnan Observatories, Chinese Academy of Sciences, Kunming 650011, People's Republic of China}
\affiliation{Key Laboratory for the Structure and Evolution of Celestial Objects, Chinese Academy of Sciences, Kunming 650011, People's Republic of China} 
\author[0000-0001-9449-9268]{Jian-Min Wang}
\affiliation{Key Laboratory for Particle Astrophysics, Institute of High Energy Physics, Chinese Academy of Sciences, 19B Yuquan Road, Beijing 100049, People's Republic of China} 
\author{Chen Hu}
\affiliation{Key Laboratory for Particle Astrophysics, Institute of High Energy Physics, Chinese Academy of Sciences, 19B Yuquan Road, Beijing 100049, People's Republic of China}
\author[0000-0001-5841-9179]{Yan-Rong Li}
\affiliation{Key Laboratory for Particle Astrophysics, Institute of High Energy Physics, Chinese Academy of Sciences, 19B Yuquan Road, Beijing 100049, People's Republic of China}
\author[0000-0002-5830-3544]{Pu Du}
\affiliation{Key Laboratory for Particle Astrophysics, Institute of High Energy Physics, Chinese Academy of Sciences, 19B Yuquan Road, Beijing 100049, People's Republic of China}
\author{Ming Xiao}
\affiliation{Key Laboratory for Particle Astrophysics, Institute of High Energy Physics, Chinese Academy of Sciences, 19B Yuquan Road, Beijing 100049, People's Republic of China}
\author[0000-0002-1530-2680]{Hai-Cheng Feng}
\affiliation{Yunnan Observatories, Chinese Academy of Sciences, Kunming 650011, People's Republic of China} 
\affiliation{Key Laboratory for the Structure and Evolution of Celestial Objects, Chinese Academy of Sciences, Kunming 650011, People's Republic of China} 
\author{Sha-Sha Li}
\affiliation{Yunnan Observatories, Chinese Academy of Sciences, Kunming 650011, People's Republic of China} 
\affiliation{Key Laboratory for the Structure and Evolution of Celestial Objects, Chinese Academy of Sciences, Kunming 650011, People's Republic of China} 
\author{Jian-Guo Wang}
\affiliation{Yunnan Observatories, Chinese Academy of Sciences, Kunming 650011, People's Republic of China} 
\affiliation{Key Laboratory for the Structure and Evolution of Celestial Objects, Chinese Academy of Sciences, Kunming 650011, People's Republic of China} 
\author[0000-0002-2419-6875]{Zhi-Xiang Zhang}
\affiliation{Department of Astronomy, Xiamen University, Xiamen, Fujian 361005, People's Republic of China}
\author{Ying-Ke Huang}
\affiliation{Multi-disciplinary Research Division, Institute of High Energy Physics, Chinese Academy of Sciences, 19B Yuquan Road, Beijing 100049, People's Republic of China}

\email{lukx@ynao.ac.cn}

\begin{abstract}
NGC 5548 is one of the active galactic nuclei (AGN) selected for our long-term spectroscopic monitoring with the Lijiang 2.4~m telescope, aiming at investigating the origin and evolution of the broad-line regions (BLRs), accurately measuring the mass of the supermassive black holes (SMBHs), and understanding structure and evolution of the AGN. 
We have performed five-season observations for NGC~5548 with the median sampling interval ranging from 1.25 to 3 days. 
The light curves of the 5100~\AA\ continuum and broad emission lines are measured 
after subtracting contamination of the host galaxy starlight. 
The time lags of the broad He~{\sc ii}, He~{\sc i}, H$\gamma$, and H$\beta$ lines with respect to the 5100~\AA\ continuum are obtained for each season and their mean time lags over the five seasons are 0.69, 4.66, 4.60, 8.43 days, respectively. 
The H$\gamma$ and H$\beta$ velocity-resolved lag profiles in the seasons of 2015, 2018, 2019, and 2021 are constructed, 
from which an ``M-shaped'' structure is found in 2015 but disappears after 2018. 
Our five-season reverberation mapping (RM) yields an averaged virial SMBH mass of $M_\bullet/10^7M_\odot=14.22$, with a small standard deviation of $1.89$. 
By combining the previous 18 RM campaigns and our five-season campaign for NGC~5548, we find that there exists a time lag of 3.5~years between the changes in the BLR size and optical luminosity. 
In addition, we also construct the BLR radius$-$luminosity relation and the virial relation for NGC~5548. 
\end{abstract}

\keywords{Active galactic nuclei (16); Supermassive black holes (1663); Reverberation mapping (2019); Time domain astronomy (2109)}

\section{Introduction} \label{sec:intro}
Reverberation mapping (RM; \citealt{Blandford1982,Peterson1993}) 
has been widely used in spectroscopic monitoring campaigns to probe kinematics of broad-line region (BLR) 
and measure masses of the accreting supermassive black hole (SMBH) in the centres of active galactic nuclei 
(AGN; e.g., \citealt{Peterson1999,Bentz2009,Denney2010,Du2018b,Lu2021a,Hu2021}). 
RM has also been employed in multi-band photometric monitoring campaigns to measure accretion disc sizes (e.g., \citealt{Zhu2018, Cackett2020, Guo2022}). 
Before 2013, RM measurements of $\sim$50 AGNs were obtained by different spectroscopic monitoring campaigns, from which 
the canonical radius$-$luminosity (i.e, $R_{\rm BLR}-L_{5100}$) relation were established (\citealt{Kaspi2000, Bentz2013}). 
However, this RM sample is heterogeneous and mainly consists of sub-Eddington AGNs. 

In the recent 10 years, based on the different samples, the canonical $R_{\rm BLR}-L_{5100}$ relation has been tested for different purposes. 
The SEAMBH (super-Eddington accreting massive black holes) project focused on spectroscopic RM of high accretion AGNs. The major finding was that the H$\beta$ time lags in SEAMBHs are significantly shorter than those in sub-Eddington AGNs (\citealt{Du2018b}) 
and accretion rate is the main driver for the shortened lags (\citealt{Du2019}). 
This finding was subsequently confirmed by the Sloan Digital Sky Survey Reverberation Mapping project, 
which also found that there are many AGNs located below the canonical $R_{\rm BLR}-L_{5100}$ relation (\citealt{Grier2017}). 
\cite{Hu2021} and \cite{Liss2021} carried out high-cadence spectroscopic monitoring on a number of PG quasars 
and found that some objects have H$\beta$ time lags shortened by almost 0.3~dex. 
To investigate the deviations of $R_{\rm BLR}-L_{5100}$ relation, 
\cite{Du2018a} monitored a sample of AGNs with complex H$\beta$ line profiles. 
\cite{Lu2019a} developed a spectroscopic monitoring project for AGNs with disc winds$/$outflows to investigate whether their BLRs originate from the disc winds or not. 
Not long after that, \cite{Matthews2020} proposed that the BLR might be from the failed accretion disc winds. 
New insights into the deviations from the canonical $R_{\rm BLR}-L_{5100}$ relation can be gained through enlarging the RM sample size and expanding the dynamic range of the sample's properties (such as luminosity, SMBH mass, and accretion rate).  
 
On the other hand, investigating BLR evolution in radius and kinematics for individual AGNs along with different luminosity (or accretion) states 
can provide a new perspective for understanding the deviations from the canonical $R_{\rm BLR}-L_{5100}$ relation. 
This can be implemented through a multi-season RM campaign for extremely variable AGNs. 
Fortunately, more and more extremely variable AGNs$/$quasars were found from archival data, 
such as changing-look AGNs$/$quasars (CL~AGN) or changing-state quasars (CSQ) and tidal disruption events 
(e.g., \citealt{Yang2018,Shu2018,Graham2020,Liu2021,Feng2021}), 
lending us good AGN samples with BLR properties at different luminosity (or accretion) states. 
With this motivation, we carried out a long-term  project using the Lijiang 2.4~m telescope to spectroscopically monitor a selected sample of  previously famous AGNs, including RM AGNs (e.g., NGC~5548), 
changing-look AGNs (e.g., Mrk~1018, Mrk~590), and candidates of SMBH binary 
(e.g., SDSS~J153636.22+044127.0, which also is a double-peaked AGN in broad emission lines). 
In this work, we report the results of NGC~5548 from our five-season RM campaign. 

NGC~5548 (14:17:59.534, +25:08:12.44, $z=0.01717$), had previously been monitored by 18 observing seasons before 2015, which underwent extreme variability in the past two decades (e.g., \citealt{Lu2016,Li2016}). 
Based on the RM measurements of NGC~5548, \cite{Lu2016} found that its BLR radius follows the varying optical luminosity, 
but with a tentative time delay of 2.35~years. They suggested that such a time delay might be related to the radiation pressure from the central accretion disc. 
Meanwhile, \cite{Li2016} and \cite{Bon2016} detected periodic variations of the continuum and double-peaked profiles of the broad H$\beta$ line in NGC~5548, implying that NGC~5548 is a possible sub-parsec SMBH binary candidate. 
Besides, a rare phenomenon called ``BLR holiday'', namely, the broad emission lines are strongly decorrelated from the AGN continuum, 
was observed in NGC~5548 in the year of 2014 (\citealt{Goad2016,Pei2017}). 
The falling corona model (\citealt{Sun2018}) and disc wind model (\citealt{Dehghanian2020}) 
were proposed to explain this anomalous behavior, however, there is not yet a consensus on the final decisive explanation. 
Taken together, it is highly worthwhile to investigate the kinematics of the BLR in NGC~5548 and its variations over a long timescale. 

NGC~5548 was therefore the highest priority target of our long-term spectroscopic monitoring campaign. 
In 2015, we conducted the first season of observations. Between 2018 and 2021, we continuously performed four seasons of spectroscopic monitoring. 
Hereafter, we refer to these five seasons as the season of 2015, 2018, 2019, 2020, and 2021, respectively. 
The season of 2015 started on January 07, 2015 and ended on August 01, 2015, 
the RM measurement of the broad H$\beta$ line as the first result has been reported by \cite{Lu2016}. 
In this work, we not only include this result for the sake of completeness, 
but also provide other RM measurements including broad H$\gamma$, He~{\sc ii}, and He~{\sc i} lines for this season. 

The paper is organized as follows. Section~\ref{sec:od} describes the observation and data reduction. 
Section~\ref{sec:data} presents data analysis, including the measurements of light curves, time lags, variability characteristics, and line widths, along with velocity-resolved RM analysis. 
Section~\ref{sec:prop} compares our RM results with the previous 18 RM measurements and investigates the BLR radius$-$luminosity relation, the virial relation, 
and the secular variation of the BLR in NGC~5548. In Seciton~\ref{sec:mass}, we estimate the virial SMBH mass of NGC 5548. 
We end with a summary of our main results in Section~\ref{sec:sum}. 
Throughout the paper, we use a cosmology with 
$H_0=72{\rm~km~s^{-1}~Mpc^{-1}}$, $\Omega_{\Lambda}=0.7$, and $\Omega_{\rm M}=0.3$.

\section{Observation and Data Reduction} \label{sec:od}
After the season of 2015, we continuously monitored NGC~5548 between 2018 and 2021. 
The spectroscopic observation settings and data reduction were similar to those in \cite{Lu2016} for NGC~5548. 
The readers are also referred to our previous works on other AGNs Mrk 79, NGC~7469, and Mrk~817 \citep{Lu2019a,Lu2021a}
for more detailed discussions on the RM experiments. 

\subsection{Spectroscopy} \label{sec:so}
The spectroscopic observations of NGC~5548 were taken using 
Yunnan Faint Object Spectrograph and Camera (YFOSC) mounted on the Lijiang 2.4~m telescope (LJT), 
which locates in the Lijiang observatory of Yunnan Observatories, Chinese Academy of Sciences. 
YFOSC is equipped with a  back-illuminated 2048$\times$2048 pixel CCD, 
with the pixel size 13.5 $\mu$m, pixel scale 0.283$''$ per pixel, and field-of-view $10'\times10'$. 
It is a versatile instrument for low-resolution spectroscopy and photometry. 
More information about the Lijiang observatory and telescope was described in \cite{Fan2015}, \cite{Wang2019}, \cite{Xin2020} and \cite{Lu2021b}. 

\begin{figure*}[htb]
\centering
\includegraphics[angle=0,width=0.95\textwidth]{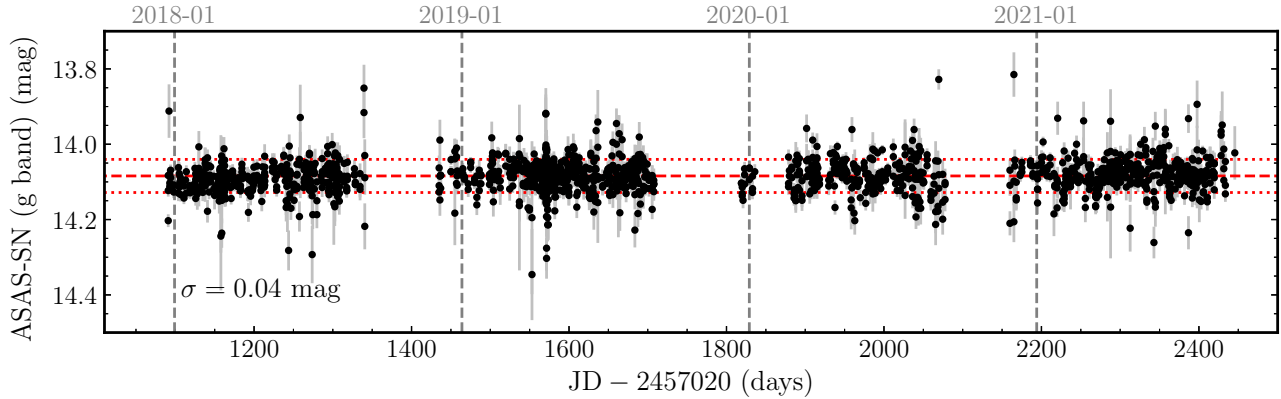}
\caption{\footnotesize
The $g$-band light curves of NGC~5548's comparison star observed by the All-Sky Automated Survey for Supernovae (ASAS-SN) 
in 2018, 2019, 2020, and 2021. The scatter of light curves is 0.04~mag (i.e., 4\%), which is comparable with {\bf its} averaged photometric uncertainty of 0.03~mag. 
The $V$-band light curves observed by ASAS-SN in 2015 are not included. 
}
\label{fig_complc}
\end{figure*}

Following the observations in the season of 2015,  
we oriented a long slit in the field of view to take spectra of NGC~5548 and a nearby non-varying comparison
star simultaneously. 
This observation method was described in detail by \citet{Maoz1990} and \citet{Kaspi2000}, 
and widely adopted by recent RM campaigns (e.g., \citealt{Du2015,Lu2021a}). 
The adopted comparison star, SDSS~J141758.82+250533.1 
(hereafter J1417), has a spectral type of G1 and $V$-band magnitude of 13.9. 
The temperature and radius of the comparison star obtained from 
our SED fitting and spectral matching are all consistent with 
the result of {\it Gaia} DR3 \citep{Gaia2021}. 
The angular distance between the comparison star and NGC~5548 is $160''$. 
During the seasons of 2018, 2019, 2020, and 2021, 
J1417 was also monitored by the project of All-Sky Automated Survey 
for Super-novae (ASAS-SN; \citealt{Shappee2014,Kochanek2017}) 
with the g band filter. 
The photometric data were downloaded from the site\footnote{http://www.astronomy.ohio-state.edu/asassn/index.shtml} and used to check the stability of J1417.
The resulting $g$-band light curve is displayed in Figure~\ref{fig_complc}, with a scatter of 0.04~mag. 
This scatter is comparable with the averaged photometric error (0.03 mag) of this light curve, 
implying that J1417's emission is stable enough so that J1417 was selected as the reference star of NGC~5548. 
Our previous works demonstrated that comparison star as a reference standard can provide a high-precision flux calibration (\citealt{Lu2019a}, see also \citealt{Hu2015}). 
In some cases, the spectra of the comparison star can be used to calibrate the telluric absorption lines of the target's spectra (\citealt{Lu2021b}). 
In light of the average seeing of the observatory site, 
we adopted a long slit with a projected width of $2.5''$. We used Grism 14 which covers the wavelength from $\sim$3600~\AA\ to 7460~\AA\ and provides a dispersion of 1.8~\AA\,~pixel$^{-1}$. 
The standard neon and helium lamps were used for wavelength calibration. 
 
Totally we obtained 315 spectroscopic observations for NGC~5548. 
In each season, the observations generally spanned from January to June. The median sampling intervals of the five seasons range from 1.25 to 3 days. 
Table~\ref{tab_obsst} is observation statistics of the five seasons. 

\begin{deluxetable}{cccccc}
\tablecolumns{6}
\tabletypesize{\scriptsize}
\setlength{\tabcolsep}{3pt}
\tablewidth{2pt}
\tablecaption{Observation statistics of NGC~5548 from the Lijiang 2.4~m telescope\label{tab_obsst}}
\tablehead{
\colhead{Season}                 &
\colhead{Dates}                  &
\colhead{Period}                   &
\colhead{$N$}                       &
\colhead{$<T>$}                   &
\colhead{$T_{\rm median}$} \\
\colhead{}                              &
\colhead{(yy/mm/dd)}            &
\colhead{(days)}                    &
\colhead{}                              &
\colhead{(days)}                    &
\colhead{(days)}                    \\
\colhead{(1)} &
\colhead{(2)} &
\colhead{(3)} &
\colhead{(4)} &
\colhead{(5)} &
\colhead{(6)} 
}
\startdata
2015 & 2015 Jan. 07$-$2015 Aug. 01 & 206 & 62 & 3.40 & 3.00 \\
2018 & 2018 Mar. 12$-$2018 Jun. 18 & 98   & 40 & 2.51 & 1.25 \\
2019 & 2018 Nov. 28$-$2019 Jun. 20 & 204 & 81 & 2.55 & 2.00 \\
2020 & 2020 Jan. 11$-$2020 Jun. 21 & 163 & 52 & 3.19 & 2.00 \\
2021 & 2020 Dec. 24$-$2021 Aug. 06 & 225 & 80 & 2.84 & 1.75 
\enddata
\tablecomments{\footnotesize
Column (1) is the season (including 2015, 2018, 2019, 2020, and 2021). 
Column (2) gives the dates of spectroscopic monitoring. 
Columns (3) and (4) are the observation period and the total number of sampling. 
Columns (5) and (6) are mean and median spectroscopic sampling intervals. \\
}
\end{deluxetable}

\begin{figure}[htb]
\centering
\includegraphics[angle=0,width=0.49\textwidth]{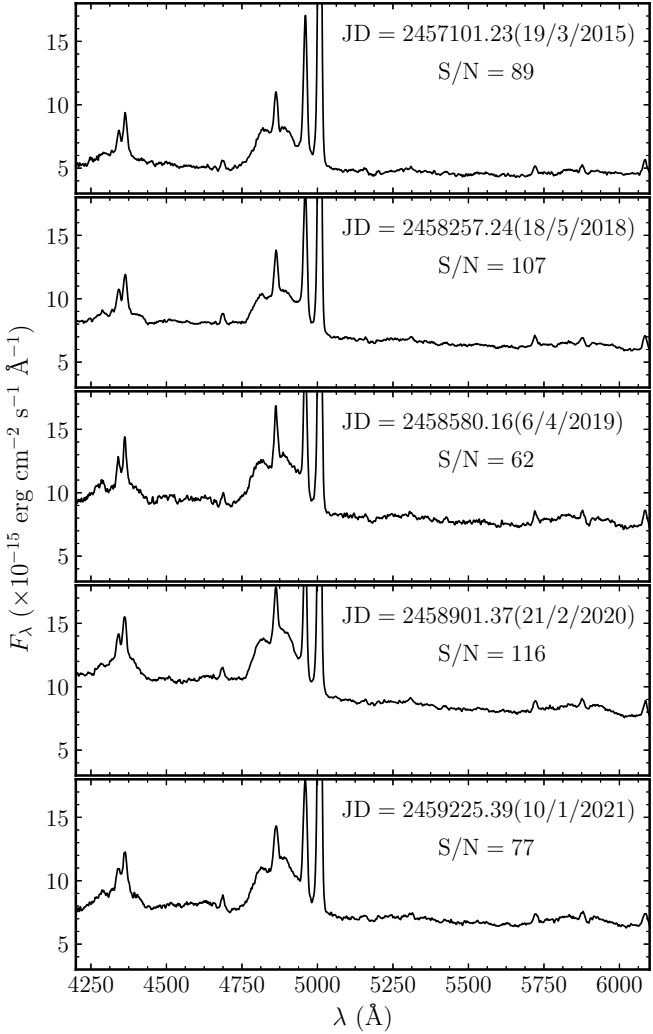}
\caption{\footnotesize
We randomly select one spectrum per season from the calibrated ones (see Section~\ref{sec:dr}) 
to showcase spectral quality. Julian Date (including calendar) and signal-to-noise (S/N) ratio at 5100~\AA\ are noted in each panel. 
}
\label{fig_ind}
\end{figure}

\begin{figure}[htb]
\centering
\includegraphics[angle=0,width=0.48\textwidth]{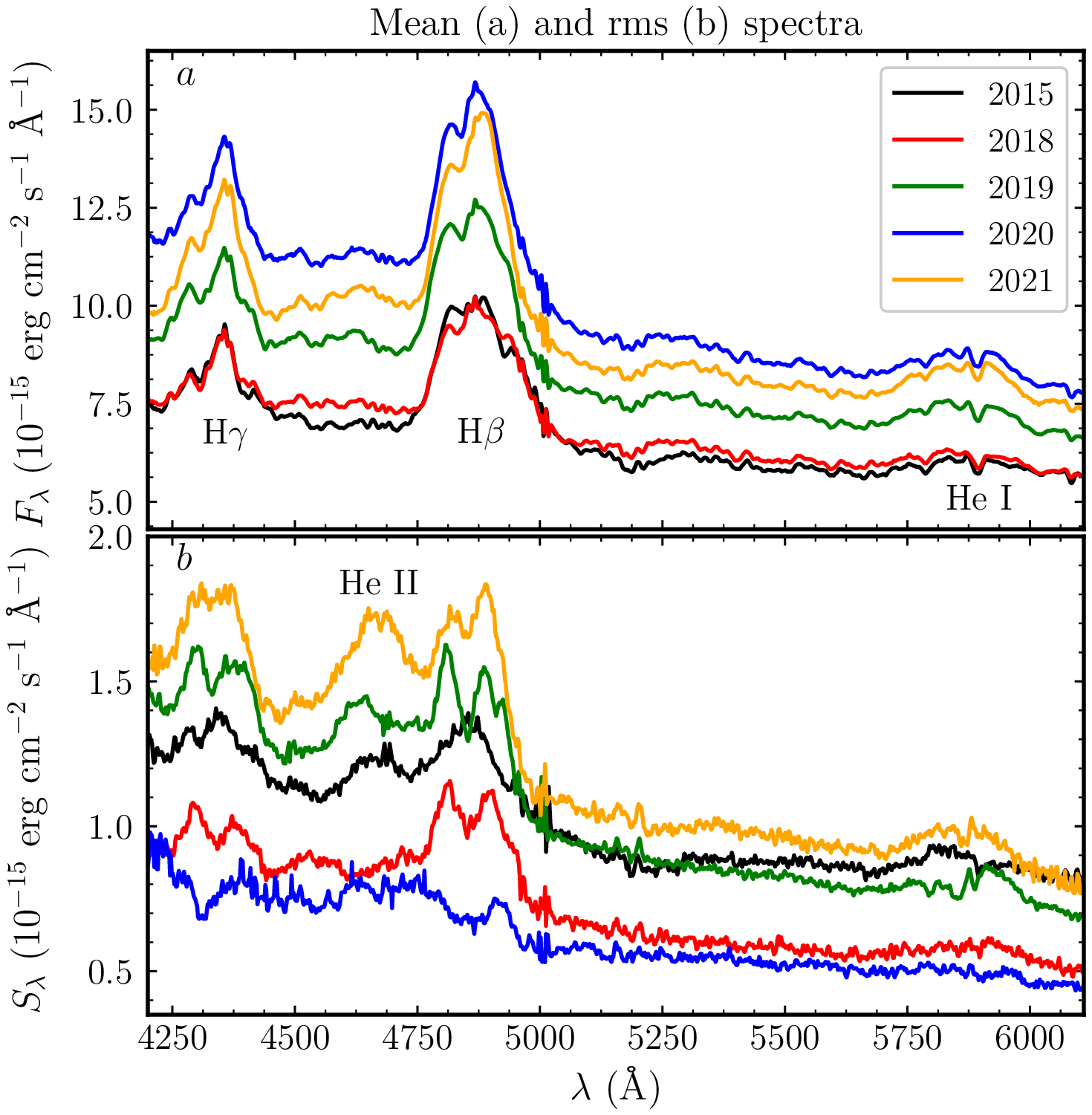}
\caption{\footnotesize
Mean and root-mean-square (rms) spectra of each season calculated from the processed spectra (see Section~\ref{sec:dr}), 
the legend in the panel~({\it a}) marks the seasons of 2015, 2018, 2019, 2020, and 2021. 
Just narrow emission lines (such as narrow H$\gamma$, He~{\sc ii}, H$\beta$, [O~{\sc iii}] and He~{\sc i}) 
were subtracted by spectral fitting and decomposition (see Section~\ref{sec:lc}) before calculation. 
}
\label{fig_meanrms}
\end{figure}

\subsection{Data Reduction}\label{sec:dr}
Two-dimensional spectra were reduced using the standard {\tt IRAF} procedures, 
which include bias subtraction, flat-field correction, wavelength calibration, and cosmic ray elimination. 
A relatively small extraction window helps to reduce the Poisson noise of sky-background 
and increase the signal-to-noise (S/N) ratio of the spectra. 
Therefore, a uniform extraction window of 20 pixels (5.7$^{\prime\prime}$) was used and 
the sky-background determined from two adjacent regions ($+7.4^{\prime\prime}\sim+14^{\prime\prime}$ 
and $-7.4^{\prime\prime}\sim-14^{\prime\prime}$) on both sides of the extraction window was subtracted. 

The scientific target NGC~5548 and its comparison star were observed simultaneously in a long slit 
with the same observing conditions (such as airmass and seeing), 
so that the spectrum of NGC~5548 can be calibrated accurately by the sensitivity function calculated from the comparison star (see \citealt{Lu2019a}). 
Following the previous works (e.g., \citealt{Du2015,Lu2016,Lu2021a}), 
we first generated the fiducial spectrum of the comparison star using data from nights with photometric conditions, 
and obtained the sensitivity function by comparing the observed spectrum of the comparison star in each exposure to the fiducial spectrum. 
Then this sensitivity function was applied to calibrate the spectrum of NGC~5548. 
For each calibrated spectrum, 
the Galactic extinction was corrected using the extinction map of \cite{Schlegel1998}, 
Wavelength shift usually caused by varying seeing and mis-centering was corrected using [O~{\sc iii}]~$\lambda5007$ line as the wavelength reference. 
Then we transformed all spectra into the rest frame for subsequent analysis. 
The averaged S/N ratio of the spectra at 5100~\AA\ over the five seasons is 77, with a standard deviation of 28. 
We randomly select one spectrum per season and display the five spectra in Figure~\ref{fig_ind} to showcase the spectral quality. 

\section{Data Analysis}\label{sec:data}
\subsection{Mean and RMS Spectra}\label{sec:meanrms}
In this section, we calculate the mean and root-mean-square (rms, i.e., variable spectrum) 
spectrum for the seasons of 2015, 2018, 2019, 2020, and 2021, respectively. 
The mean spectrum is defined as (\citealt{Peterson2004}) 
\begin{equation}
F_{\lambda}=\frac{1}{N}\sum_{i=1}^NF_i(\lambda), 
\label{eq_msp}
\end{equation}
where $F_i(\lambda)$ is the $i$th spectrum and $N$ is the total number of spectra of each season (see Table~\ref{tab_obsst}). 
The rms spectrum is defined as 
\begin{equation}
S_{\lambda}=\left\{\frac{1}{N-1}\sum_{i=1}^N\left[F_i(\lambda)-F(\lambda)\right]^2\right\}^{1/2}. 
\label{eq_rsp}
\end{equation}
At first, the narrow-line components of the calibrated spectra are eliminated during the spectral fitting and decomposition (see Section~\ref{sec:lc}). 
Because the narrow emission lines usually have apparent variations caused by the varying seeing (e.g., [O~{\sc iii}] doublets, see Figure~11 of \citealt{Lu2019a}), 
this treatment is helpful in single out the variable spectrum of the broad emission lines. 
Then the calibrated spectra without narrow lines are used to calculate the mean and rms spectra. 
The results are displayed in Figure~\ref{fig_meanrms}. 
We can find that the optical radiation of NGC~5548 reaches the maximum in the season of 2020, 
but its variability gets the lowest. 

\begin{figure}[htb]
\centering
\includegraphics[angle=0,width=0.49\textwidth]{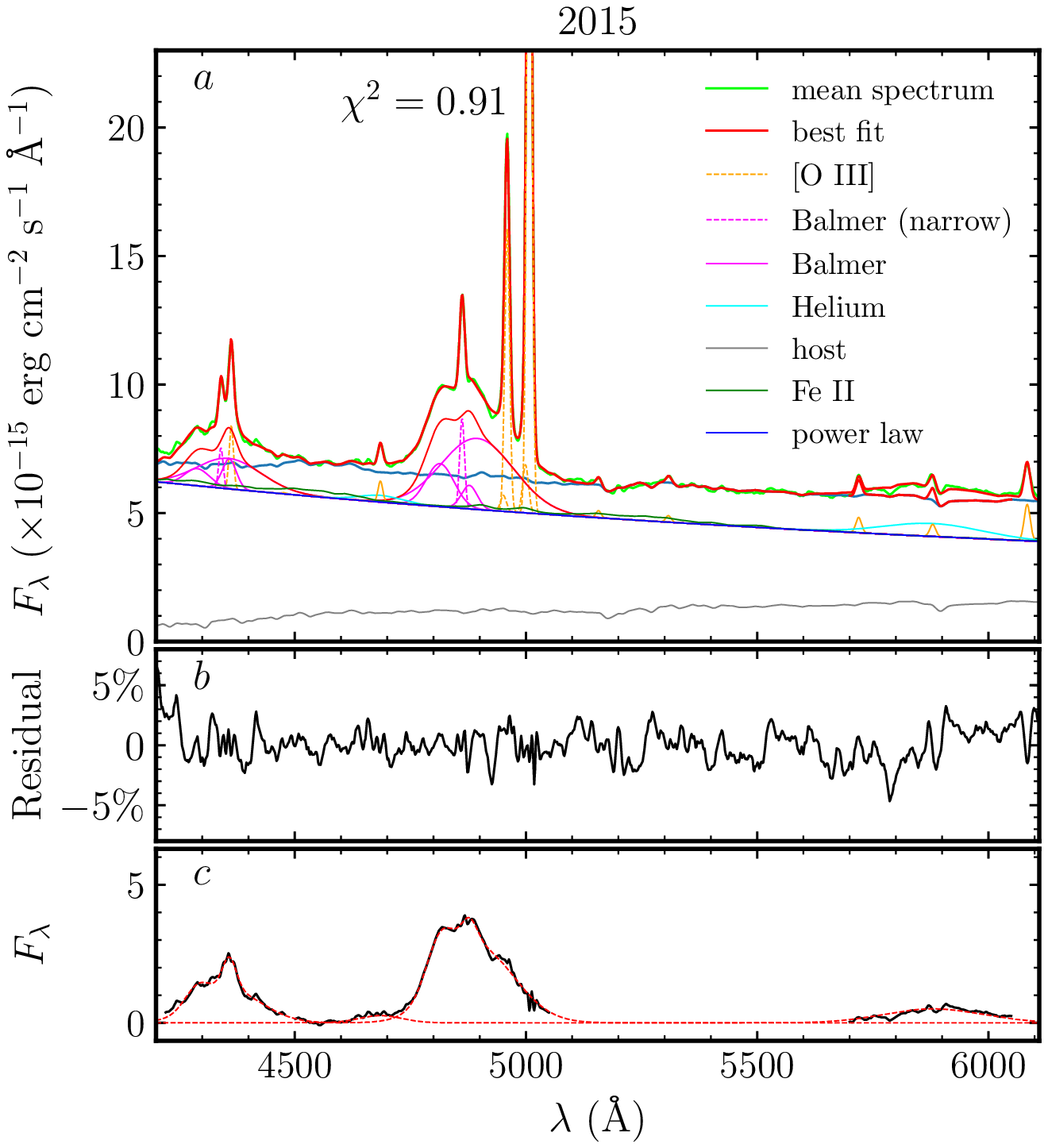}\\
\caption{\footnotesize
Multicomponent fitting and decomposition of mean spectrum calculated from the season of 2015. 
Panel~({\it a}) shows the details of spectral fitting and decomposition, 
where the mean spectrum shows in lime, 
and the fitting components are noted in the legend. 
Panel~({\it b}) shows the fitting residuals in percentage. 
Panel~({\it c}) shows net broad H$\gamma$, He~{\sc ii}, H$\beta$, and He~{\sc i} lines, 
where the fitted broad-line profiles (models) are in red dashed lines, 
and the broad-line profiles obtained by subtracting other fitted components from the spectrum are shown in black. 
}
\label{fig_fit1}
\end{figure}
\begin{figure*}[htb]
\centering
\includegraphics[angle=0,width=0.49\textwidth]{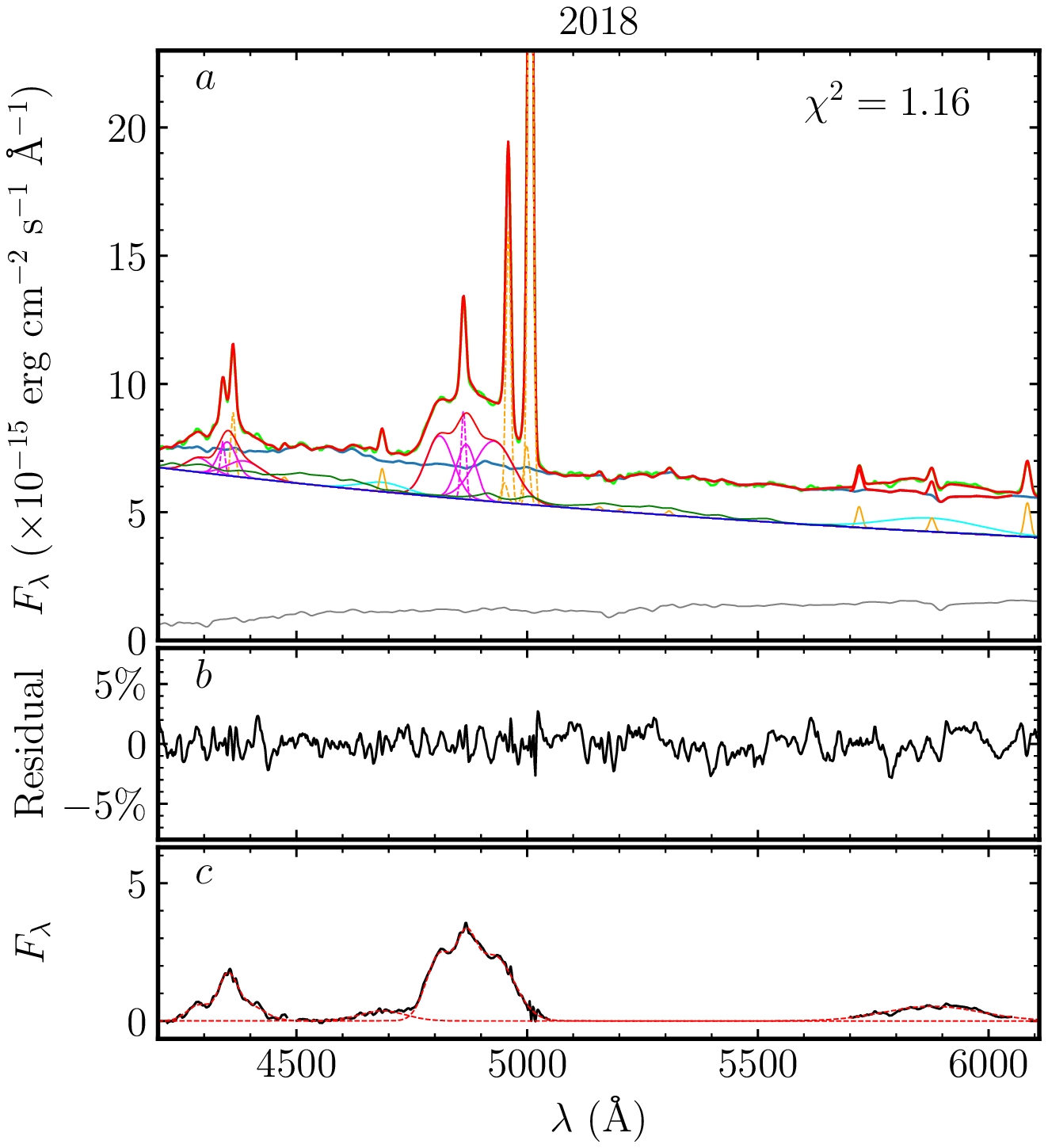}
\includegraphics[angle=0,width=0.49\textwidth]{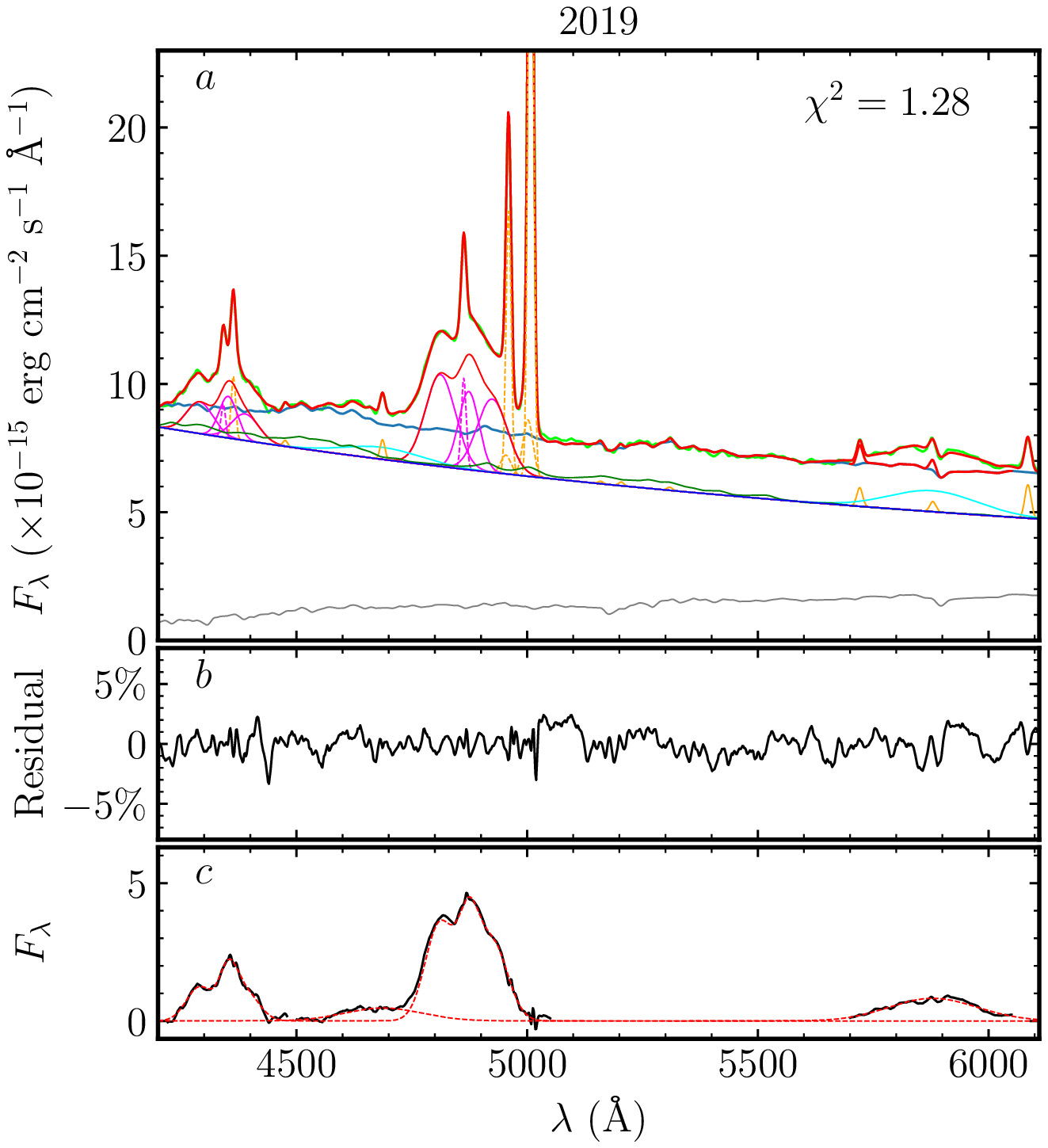}\\
\includegraphics[angle=0,width=0.49\textwidth]{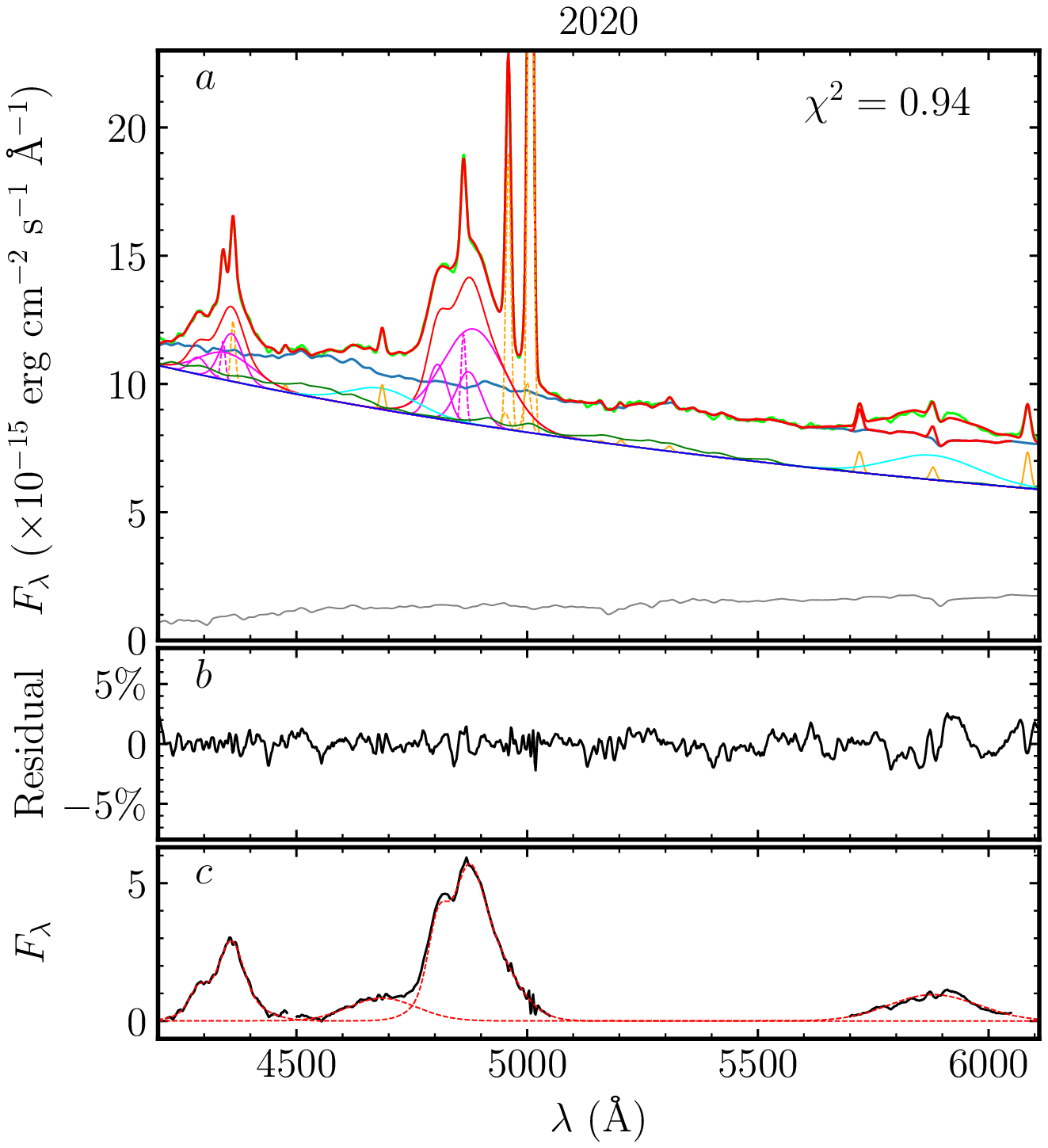}
\includegraphics[angle=0,width=0.49\textwidth]{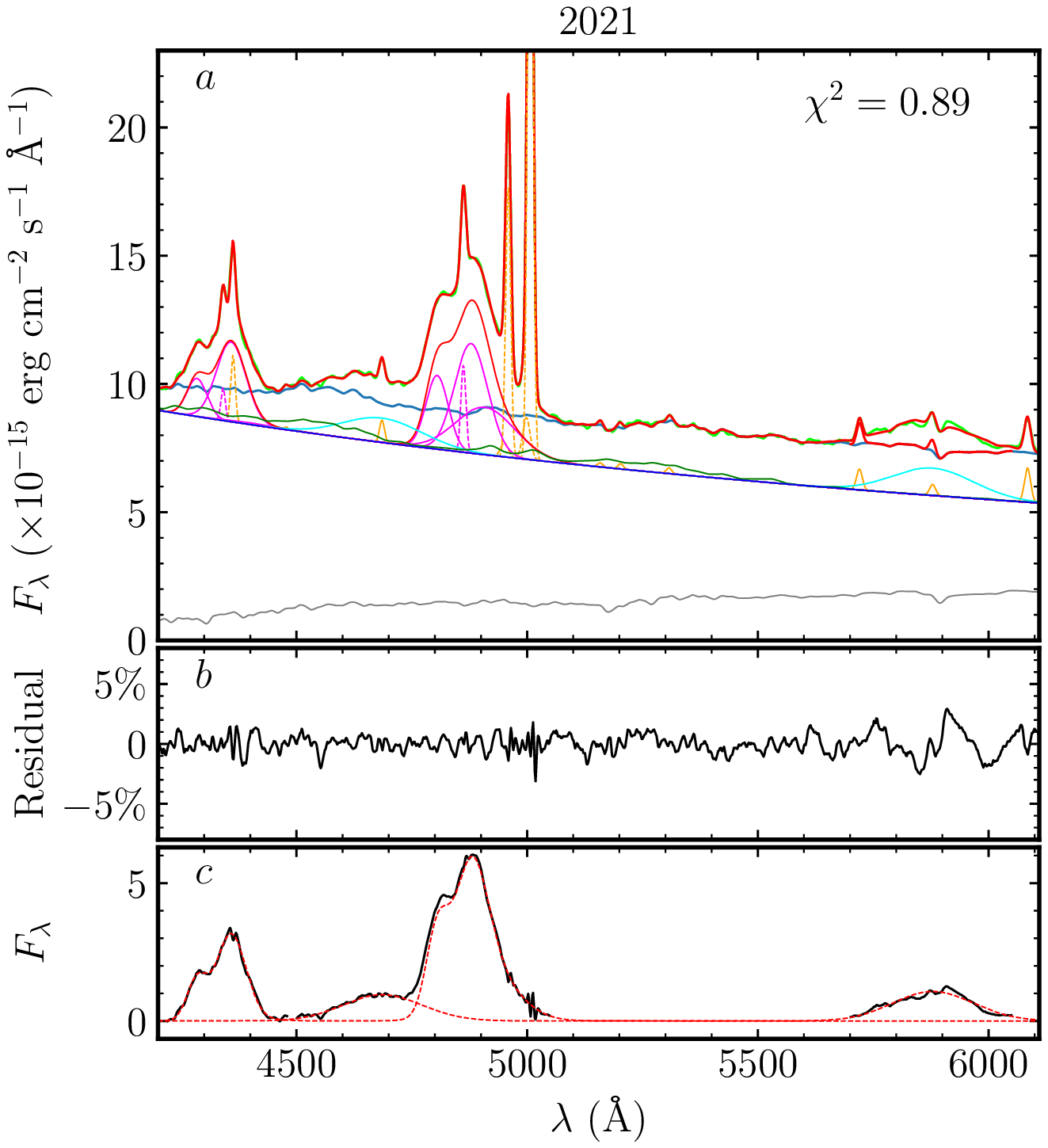}
\caption{\footnotesize
Same as Figure~\ref{fig_fit1}, but for the seasons of 2018, 2019, 2020 and 2021, respectively. 
}
\label{fig_fit2}
\end{figure*}

\subsection{Light Curves}\label{sec:lc}
In order to eliminate other blended components from the spectrum, 
the scheme of spectral fitting and decomposition is widely used in spectral analysis of AGN 
(e.g., \citealt{Hu2008,Bian2010,Dong2011,Barth2013,Guo2014,Barth2015,Lu2019c}). 
Especially in the RM campaign of local AGN, the host-galaxy starlight not only dilutes the variability of the AGN continuum and broad emission lines, 
but also usually introduces additional random noise due to nightly variations in seeing 
and target centering within the slit (\citealt{Lu2019a}, see also \citealt{Hu2015}). 
Therefore we prefer to measure the fluxes of the AGN continuum at 5100~\AA\ and 
broad H$\gamma$, He~{\sc ii}, H$\beta$, and He~{\sc i} lines using spectral fitting 
and decomposition to isolate the different components of variable spectra. 

Following the previous works (e.g., \citealt{Hu2015,Lu2021a}), 
we fit and decompose the calibrated spectra (Section~\ref{sec:dr}) at first, 
then measure the fluxes of the AGN continuum at 5100~\AA\ and broad emission lines 
from each best-fitted component, respectively. 
Our fit is based on the MPFIT package \citep{Markwardt2009}, 
which performs $\chi^{2}$-minimization using the Levenberg$-$Marquardt technique. 
The fitting window is set to be 4200$-$6110~\AA~(rest-frame). The spectrum portion at a wavelength longer than $\sim$6200~\AA\ is contaminated 
by the second-order spectrum and therefore is excluded. 
The following components are included in our fitting: 
(1) a power law ($f_{\lambda}\propto\lambda^{\alpha}$, $\alpha$ is the spectral index) 
for the AGN continuum. 
(2) An iron template from \cite{Boroson1992} for the iron multiplets. 
In practice, several iron templates were suggested for fitting the optical or ultraviolet spectrum 
of AGNs (e.g., \citealt{Boroson1992,Veron2004,Kova2010,Park2022}), 
NGC~5548 has relatively weak iron multiplets and we find that our fitting is not sensitive to the specific iron template. 
(3) The host-galaxy starlight modeled by the stellar template with an age of 11 Gyr 
and metallicity of Z = 0.05 from \cite{Bruzual2003}. 
In the individual and mean spectrum of NGC~5548, 
some stellar absorption lines are presented near the wavelength regions 
of 5876~\AA\ and 5180~\AA, which can give some limitations on the amount of the host-galaxy starlight. 
(4) Three Gaussian functions for the broad H$\beta$ and H$\gamma$ lines, respectively. 
(5) Two double Gaussians for the [O~{\sc iii}] doublets $\lambda5007/\lambda$4959. 
(6) A Gaussian for the narrow H$\beta$ line. 
(7) A set of single Gaussians for fitting other narrow emission lines. 
These components are fitted simultaneously over the fitting window. 

As the previous fitting steps (e.g., \citealt{Hu2015,Lu2021a}), 
we first fit the mean spectrum of each season and then fit the individual spectrum. 
During the fitting of the mean spectrum, 
the flux ratio of the [O~{\sc iii}] doublets is fixed to the theoretical value of 3; 
the shift of the weak broad He~{\sc ii}~$\lambda$4686 line is fixed; 
the narrow components of the [O~{\sc iii}] doublets and all the other narrow lines 
are tied with the same velocity and shift; the rest of the fitting parameters are allowed to vary. 
Meanwhile, different host-galaxy templates from \cite{Bruzual2003} are considered, however, 
the template with 11 Gyr age and metallicity Z = 0.05 gives 
a rational fit to the stellar absorption lines and the spectral index of 
AGN continuum ($\sim\lambda^{-1.5}$). Hence this template is adopted throughout. 
Figures~\ref{fig_fit1} and \ref{fig_fit2} display the results of spectral fitting 
and decomposition of the mean spectra of five seasons. 
For each season, the panel ({\it a}) shows the details of spectral fitting and decomposition, along with the reduced-$\chi^{2}$. 
The panel ({\it b}) shows the fitting residuals in percentage. 
The panel ({\it c}) shows the net broad H$\gamma$, He~{\sc ii}, H$\beta$, and He~{\sc i} lines, 
where the best-fitted broad lines are in red dashed line and the total broad-line profiles obtained by subtracting other fitted components are in black. 
From which we can inspect the general variations (including the shape of profile and flux) 
of broad lines. During the fitting of the individual spectrum, 
(1) the spectral index ($\alpha$) is fixed to the value fitted from the corresponding mean spectrum, 
since there is some degeneracy between the power-law continuum and the host galaxy (especially for spectra with a low signal-to-noise ratio), 
so decomposing host-galaxy starlight from the individual spectrum benefits from the fixed spectral index (e.g., see \citealt{Hu2015}). 
To check the impact of $\alpha$-fixed and $\alpha$-freed fitting on the results, 
we take a comparison between the $\alpha$-fixed and $\alpha$-freed light curves (including cross-correlation analysis) and find that both results are consistent; 
(2) Narrow lines are also tied with the same velocity and shift; 
(3) In addition, because the strength of the broad He~{\sc ii} line is weak, 
its line width is fixed to the best value obtained from the fitting of the mean spectrum. 
For each fitting, we calculate the reduced-$\chi^{2}$. 
The reduced-$\chi^{2}$ distribution has a median value of 1.21. %1.27, 0.58
The net broad lines of the individual spectra are constructed in a similar way to the mean spectrum. 

The fluxes of the AGN continuum ($F_{\rm 5100}$) are measured from 
the best-fitted power-law component at 5100~\AA, and the fluxes of the broad H$\gamma$, 
He~{\sc ii}, H$\beta$, and He~{\sc i} lines are measured through integration over 
the best-fitted profile. 
The light curves of these components along with the uncertainties including Poisson errors and systematic errors 
are tabulated in Table~\ref{tab_lc},  
and partially displayed in Figure~\ref{fig_map2015} (left panels) for the season of 2015, 
Figure~\ref{fig_map2018} for the season of 2018, Figure~\ref{fig_map2019} for the season of 2019, 
Figure~\ref{fig_map2020} for the season of 2020, and Figure~\ref{fig_map2021} for the season of 2021. 
The smoothing light curves of the AGN continuum at 5100~\AA\ prove
that the multicomponent coupled spectra are well decomposed. In the season of 2020, 
the light curves of the broad H$\gamma$ and H~{\sc i} lines are not displayed in Figure~\ref{fig_map2020}, 
because we did not detect the credible time lag of both light curves with respect to the varying AGN continuum, 
owing to the very low cross-correlation coefficient (see Section~\ref{sec:lag}). 

To check the quality of the spectral calibration, the photometric light curves of NGC~5548 are employed and displayed in the top panels of Figures~\ref{fig_map2015}$-$\ref{fig_map2021}, 
The $V$-band light curves  displayed in Figure~\ref{fig_map2015} were observed by the Lijiang 2.4~m telescope 
during the spectroscopic monitoring periods (see also \citealt{Lu2016}). 
The $g$-band light curves  displayed in Figures~\ref{fig_map2018}$-$\ref{fig_map2021} 
(covering the seasons of 2018, 2019, 2020, and 2021) 
are compiled from the archives of ASAS-SN (\citealt{Shappee2014,Kochanek2017}). 
By a simple inspection, we can find that the spectroscopic and photometric light curves have very similar variation structures for each season, 
implying that the spectra are well calibrated. 
In addition, we measure the fluxes of [O~{\sc iii}]~$\lambda5007$ ($F_{\rm [O~{\sc III}]}$) 
from the best-fitted profile for the whole campaign, yielding a measurement scatter of $\sim$3.5\%. 
Assuming [O~{\sc iii}]~$\lambda5007$ emission is constant, the scatter of $F_{\rm [O~{\sc III}]}$ can roughly represent the spectroscopic calibration precision of the whole campaign. 
However, in a previous RM campaign, \cite{Lu2019a} found that the varying observing conditions (e.g., seeing and mis-centering) 
will give rise to the apparent variation in flux of the extended components 
(e.g., $F_{\rm [O~{\sc III}]}$ and $F_{\rm gal}$, see Figure 11 and 12 of \citealt{Lu2019a}, also see \citealt{Hu2015}). 
Consequently, it is certain that our spectroscopic calibration precision for the whole campaign should be better than 3.5\%. 

\subsection{Variability Characteristics}\label{sec:vc}
\begin{figure*}[htb]
\centering
\includegraphics[angle=0,width=0.98\textwidth]{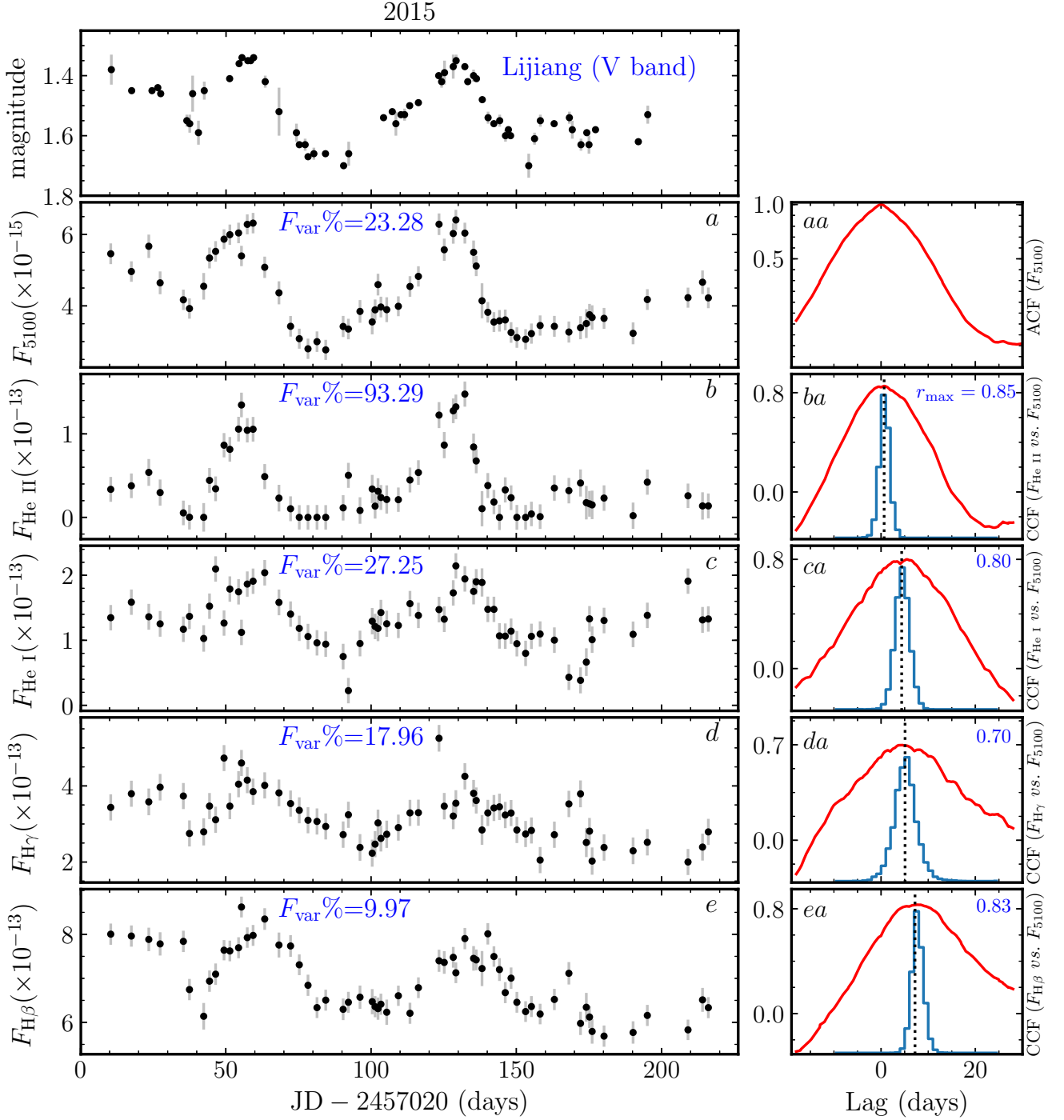}
\caption{\footnotesize
Light curves and the results of cross-correlation analysis for the season of 2015. 
The top panel shows the photometric light curve of NGC~5548 (instrument magnitude of $V$-band) 
obtained from the Lijiang~2.4~m telescope during the spectroscopic monitoring periods, which is used to check the quality of the spectral calibration. 
The left panels ({\it a-e}) are the light curves of the AGN continuum at 5100~\AA~and the broad He~{\sc ii}, He~{\sc i}, H$\gamma$, and H$\beta$ lines, 
the contamination of the host galaxy on this data has been eliminated by spectral decomposition. 
The right panels ({\it aa-ea}) are corresponding to the ACF of the continuum and the CCF between the broad-line light curves ({\it b-e}) and the continuum variation ({\it a}), the histogram in blue is the cross-correlation centroid distribution (CCCD). 
We note the variability amplitude of $F_{\rm var}\%$ in panels~({\it b-e}), 
and the maximum cross-correlation coefficient of $r_{\max}$ in CCF panels~({\it ba-ea}). 
The measured time lag is marked by the vertical dotted lines in panels~({\it ba-ea}). 
The units of $F_{\rm 5100}$ and emission lines are ${\rm erg~s^{-1}~cm^{-2}~\AA^{-1}}$ and ${\rm erg~s^{-1}~cm^{-2}}$, respectively. 
}
\label{fig_map2015}
\end{figure*}

\begin{figure*}[htb]
\centering
\includegraphics[angle=0,width=0.98\textwidth]{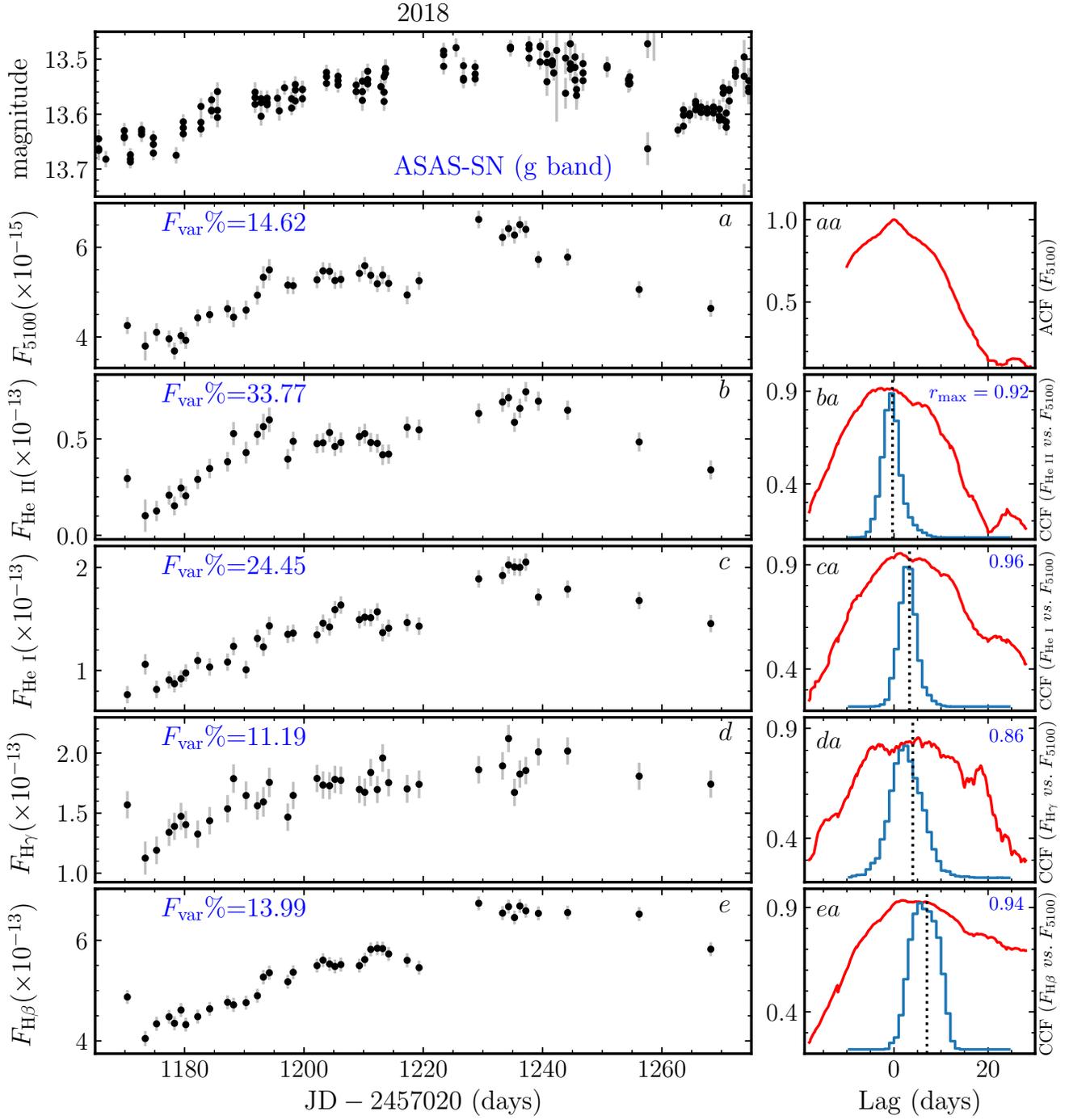}
\caption{\footnotesize
Same as Figure~\ref{fig_map2015}, but for the season of 2018, the top panel shows the photometric light curve of NGC~5548 ($g$-band) 
obtained from the ASAS-SN during the spectroscopic monitoring periods. 
}
\label{fig_map2018}
\end{figure*}

\begin{figure*}[htb]
\centering
\includegraphics[angle=0,width=0.98\textwidth]{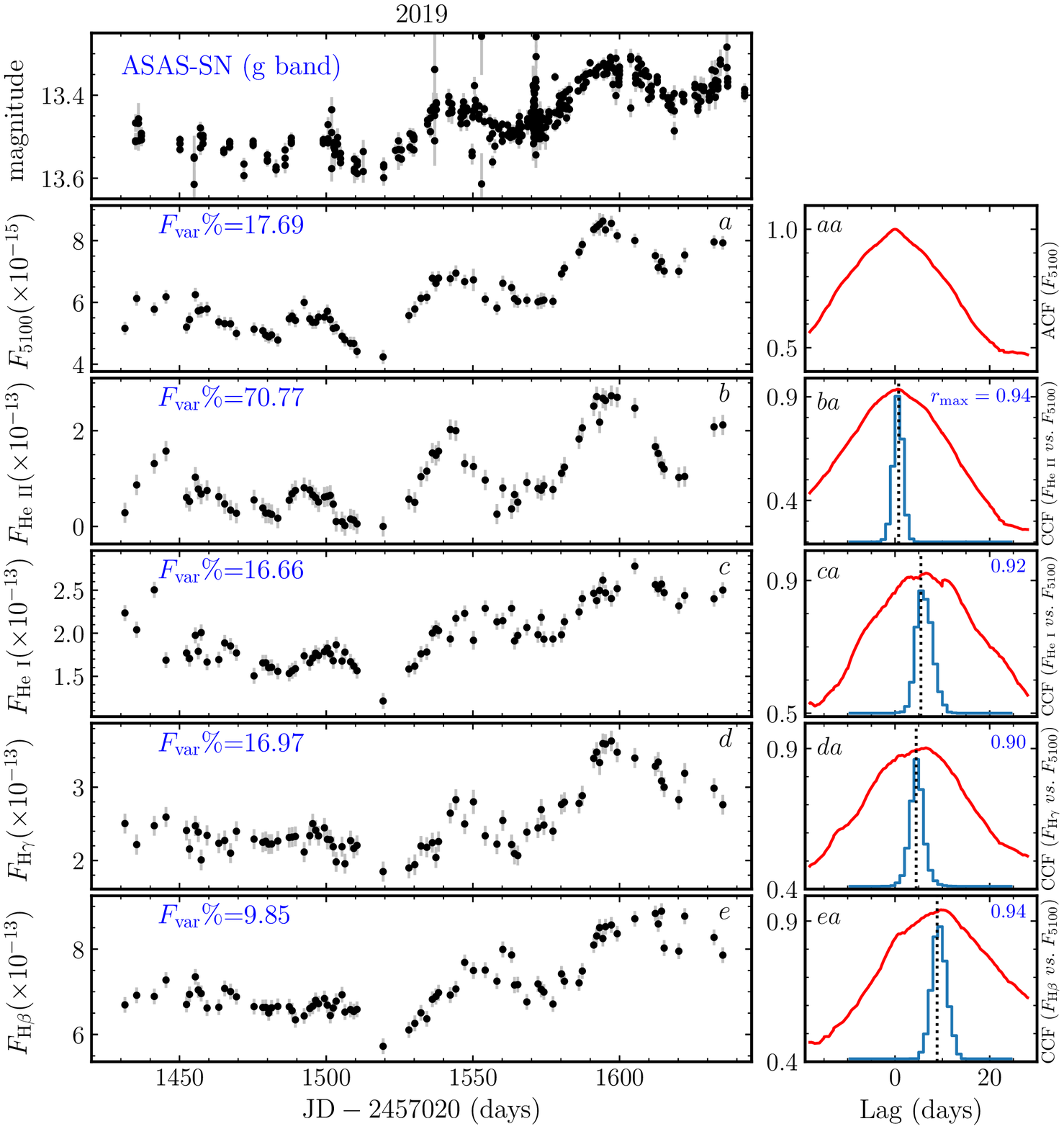}
\caption{\footnotesize
Same as Figure~\ref{fig_map2018}, but for the season of 2019. 
}
\label{fig_map2019}
\end{figure*}

\begin{figure*}[htb]
\centering
\includegraphics[angle=0,width=0.98\textwidth]{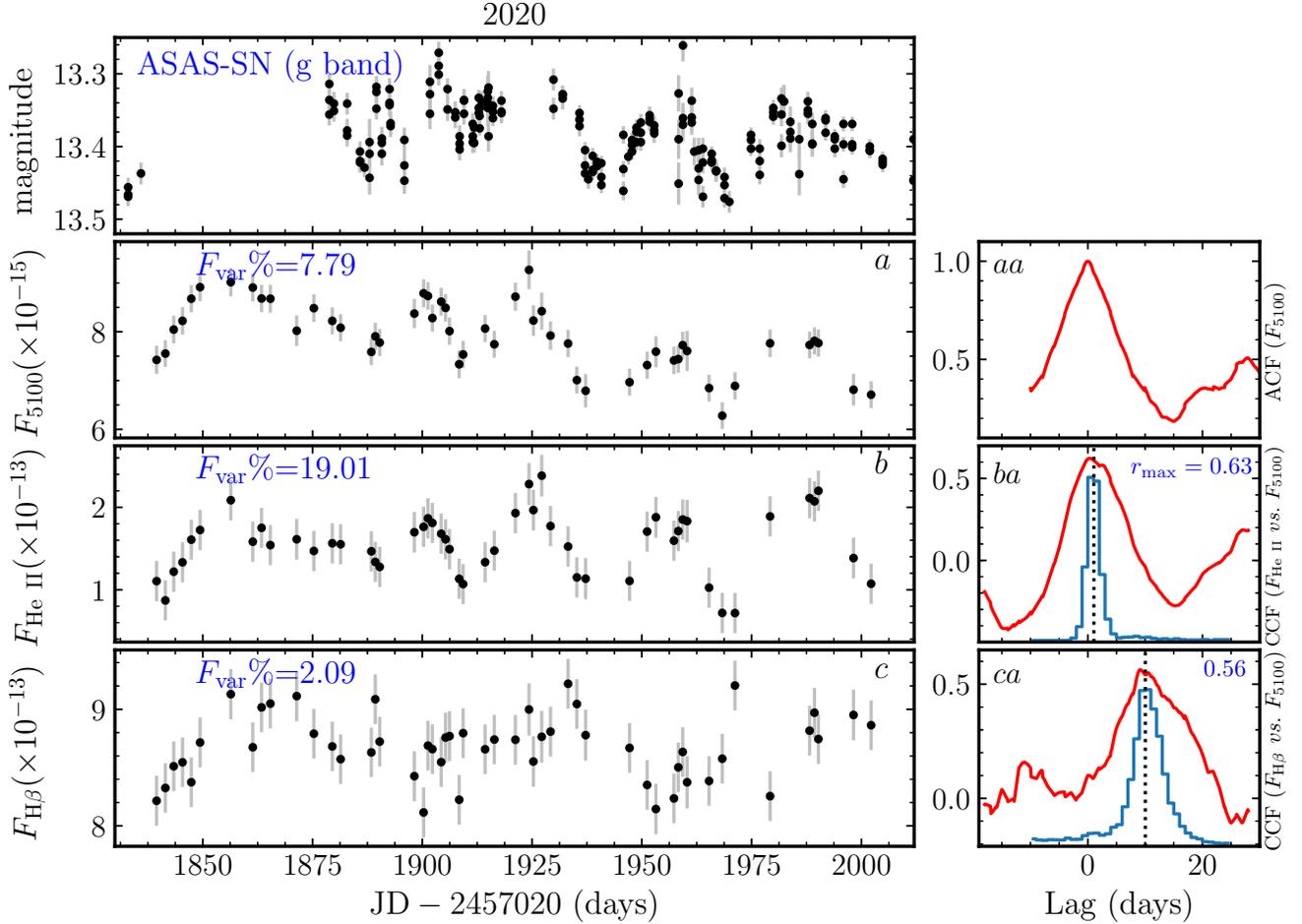}
\caption{\footnotesize
Same as Figure~\ref{fig_map2018}, but for the season of 2020. 
Because the maximum cross-correlation coefficients between the light curves of broad He~{\sc i} (and H$\gamma$ lines) and the varying 5100~\AA\ continuum 
are far less than 0.5, the relevant results are not displayed in this figure. 
}
\label{fig_map2020}
\end{figure*}

\begin{figure*}[htb]
\centering
\includegraphics[angle=0,width=0.98\textwidth]{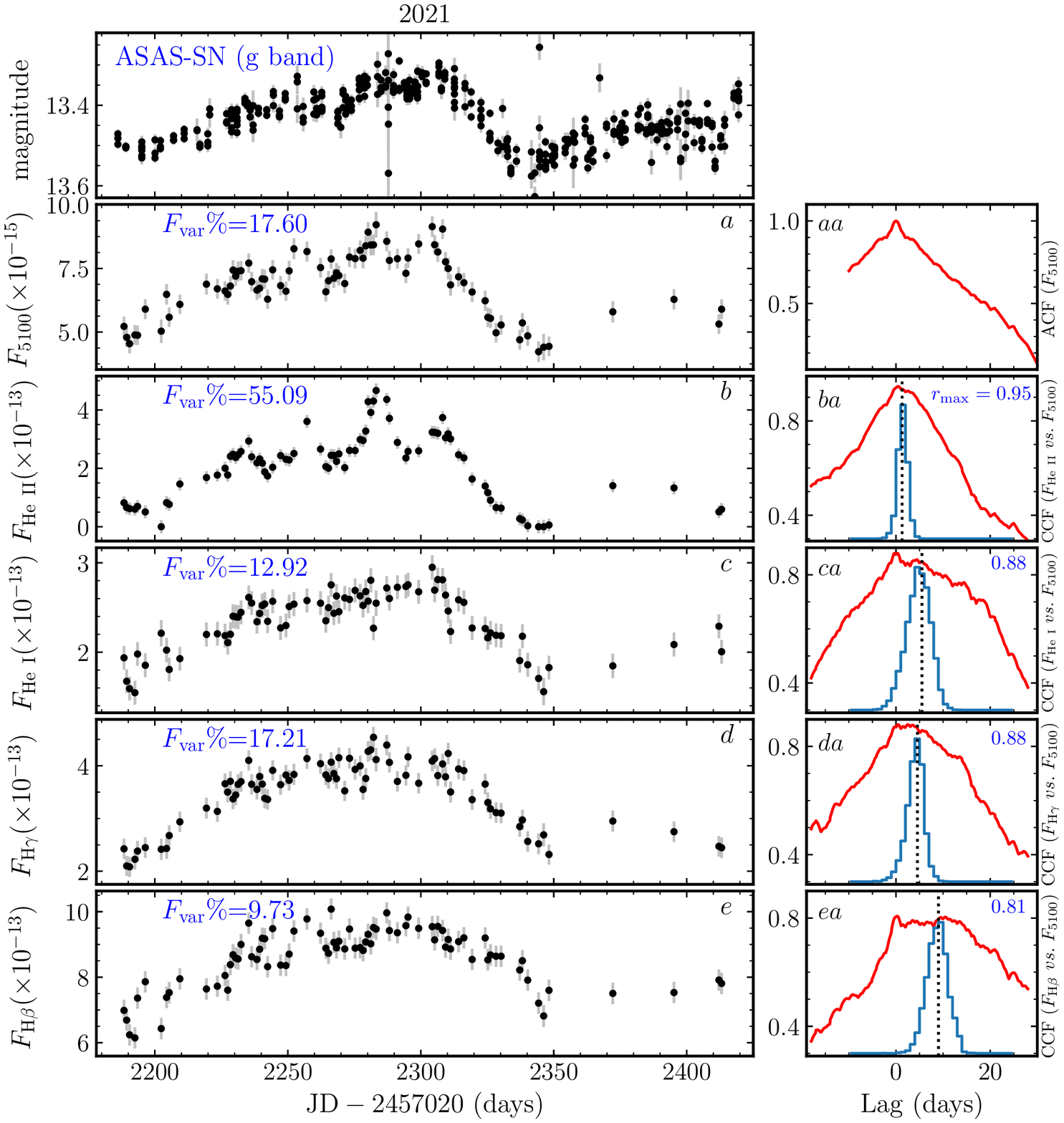}
\caption{\footnotesize
Same as Figure~\ref{fig_map2018}, but for the season of 2021. 
}
\label{fig_map2021}
\end{figure*}

We calculate the mean flux $\langle F\rangle$, standard deviation $\sigma_{\rm LC}$,  and
variability amplitude $F_{\rm var}$ (\citealt{Rodriguez-Pascual1997}) and its uncertainty of $\sigma_{F_{\rm var}}$ (\citealt{Edelson2002}) for all light curves. 
The variability amplitude $F_{\rm var}$ is defined as 
\begin{equation}
F_{\rm var}=\frac{\left(\sigma^2-\Delta^2\right)^{1/2}}{\langle F\rangle}, 
\label{eq_fvar}
\end{equation}
where $\langle F\rangle$ is the mean flux, $\sigma^2$ is the variance, and $\Delta^2$ is 
the mean square error. The uncertainty $F_{\rm var}$ is given as 
\begin{equation}
\sigma_{_{F_{\rm var}}} = \frac {1} {F_{\rm var}} \left(\frac {1}{2 N}\right)^{1/2} \frac {\sigma^2}{\langle F\rangle^2}, 
\end{equation}
where $N$ is the total epochs of each season. 
The results are listed in Table~\ref{tab_lcst} and
$F_{\rm var}$ is also marked in Figures~\ref{fig_map2015}$-$\ref{fig_map2021}. 

We find that NGC~5548 has the lowest variability in 2020 when, however, the optical radiation 
reaches the maximum. This is consistent with the finding from the mean and rms spectra 
(see Section~\ref{sec:meanrms}). This variability characteristic can be explained
by the fact that variability amplitude decreases with increasing accretion rate 
(e.g., \citealt{Lu2019b}) if the maximum optical radiation corresponds to 
the maximum accretion rate. In Section~\ref{sec:mass}, we indeed find that NGC~5548 does have the 
highest accretion rate in the season of 2020 (see Table~\ref{tab_sum}). 

\begin{deluxetable*}{cccccc}
\tablecolumns{6}
\tabletypesize{\scriptsize}
\setlength{\tabcolsep}{4pt}
\tablewidth{4pt}
\tablecaption{Light curves\label{tab_lc}}
\tablehead{
\colhead{JD}                     &
\colhead{$F_{\rm 5100}$}          &
\colhead{$F_{\rm He~II}$}         &
\colhead{$F_{\rm He~I}$}          &
\colhead{$F_{\rm H\gamma}$}  &
\colhead{$F_{\rm H\beta}$}       
}
\startdata
\multicolumn{6}{c}{The season of 2015}\\ \cline{1-6}
2457030.4 &$ 5.459\pm0.290 $&$ 0.336\pm0.149 $&$ 1.348\pm0.195 $&$ 3.436\pm0.343 $&$ 8.005\pm0.241$ \\
2457037.5 &$ 4.962\pm0.286 $&$ 0.378\pm0.148 $&$ 1.586\pm0.195 $&$ 3.795\pm0.342 $&$ 7.962\pm0.241$ \\ 
2457043.5 &$ 5.669\pm0.327 $&$ 0.538\pm0.161 $&$ 1.362\pm0.199 $&$ 3.581\pm0.351 $&$ 7.885\pm0.267$ \\
\hline
\multicolumn{6}{c}{The season of 2018}\\ \cline{1-6}
2458190.4 &$ 4.257\pm0.190 $&$ 0.295\pm0.050 $&$ 0.766\pm0.084 $&$ 1.569\pm0.114 $&$ 4.873\pm0.138$ \\
2458193.4 &$ 3.797\pm0.321 $&$ 0.101\pm0.084 $&$ 1.060\pm0.100 $&$ 1.125\pm0.138 $&$ 4.048\pm0.151$ \\
2458195.3 &$ 4.106\pm0.200 $&$ 0.126\pm0.053 $&$ 0.817\pm0.085 $&$ 1.190\pm0.115 $&$ 4.340\pm0.139$ \\
\hline
\multicolumn{6}{c}{The season of 2019}\\ \cline{1-6}
2458451.4 &$ 5.159\pm0.214 $&$ 0.285\pm0.209 $&$ 2.236\pm0.092 $&$ 2.504\pm0.134 $&$ 6.695\pm0.183$ \\
2458455.4 &$ 6.126\pm0.233 $&$ 0.867\pm0.211 $&$ 2.041\pm0.094 $&$ 2.217\pm0.137 $&$ 6.918\pm0.184$ \\
2458461.4 &$ 5.777\pm0.219 $&$ 1.313\pm0.210 $&$ 2.505\pm0.093 $&$ 2.475\pm0.135 $&$ 6.892\pm0.183$ \\
\hline
\multicolumn{6}{c}{The season of 2020}\\ \cline{1-6}
2458859.5 &$ 7.424\pm0.290 $&$ 1.104\pm0.245 $&$ 1.948\pm0.114 $&$ 2.927\pm0.160 $&$ 8.216\pm0.214$ \\
2458861.5 &$ 7.554\pm0.274 $&$ 0.870\pm0.243 $&$ 2.202\pm0.113 $&$ 2.722\pm0.156 $&$ 8.325\pm0.213$ \\
2458863.4 &$ 8.046\pm0.284 $&$ 1.219\pm0.244 $&$ 2.219\pm0.114 $&$ 3.007\pm0.158 $&$ 8.512\pm0.213$ \\
\hline
\multicolumn{6}{c}{The season of 2021}\\ \cline{1-6}
2459208.4 &$ 5.223\pm0.384 $&$ 0.820\pm0.217 $&$ 1.937\pm0.133 $&$ 2.425\pm0.192 $&$ 6.987\pm0.325$ \\
2459209.4 &$ 4.791\pm0.376 $&$ 0.658\pm0.216 $&$ 1.675\pm0.132 $&$ 2.101\pm0.191 $&$ 6.688\pm0.324$ \\
2459210.5 &$ 4.543\pm0.379 $&$ 0.619\pm0.216 $&$ 1.592\pm0.132 $&$ 2.083\pm0.191 $&$ 6.242\pm0.324$
\enddata
\tablecomments{\footnotesize
The 5100~\AA\ continuum flux is in units of ${\rm 10^{-15}~erg~s^{-1}~cm^{-2}~\AA^{-1}}$, 
all broad H$\gamma$, He~{\sc ii}, H$\beta$, and He~{\sc i} lines are in units of ${\rm 10^{-13}~erg~s^{-1}~cm^{-2}}$. 
The contamination of the host galaxy are eliminated. \\
(This table is available in its entirety in machine-readable form.) 
}
\end{deluxetable*}

\begin{deluxetable}{llccc}
\tablecolumns{5}
\tabletypesize{\scriptsize}
\setlength{\tabcolsep}{4pt}
\tablewidth{4pt}
\tablecaption{Light curve statistics\label{tab_lcst}}
\tablehead{
\colhead{Season}                 &
\colhead{Light curves}       &
\colhead{Mean flux}           &
\colhead{$\sigma_{\rm LC}$}       &
\colhead{$F_{\rm var}$ (\%)}
}
\startdata
2015& $F_{5100}$       & $4.34\pm0.30$ & 1.05 & $23.28\pm2.28$\\ 
    & $F_{\rm He~II}$  & $0.39\pm0.15$ & 0.39 & $93.29\pm9.87$\\ 
    & $F_{\rm He~I}$   & $1.31\pm0.20$ & 0.41 & $27.25\pm3.19$\\ 
    & $F_{\rm H\gamma}$& $3.21\pm0.35$ & 0.67 & $17.96\pm2.20$\\ 
    & $F_{\rm H\beta}$ & $6.95\pm0.25$ & 0.73 & $ 9.97\pm1.01$\\  \hline
2018& $F_{5100}$       & $5.13\pm0.20$ & 0.77 & $14.62\pm1.75$\\ 
    & $F_{\rm He~II}$  & $0.46\pm0.05$ & 0.16 & $33.77\pm4.21$\\ 
    & $F_{\rm He~I}$   & $1.41\pm0.09$ & 0.35 & $24.45\pm2.90$\\ 
    & $F_{\rm H\gamma}$& $1.67\pm0.12$ & 0.22 & $11.19\pm1.72$\\ 
    & $F_{\rm H\beta}$ & $5.46\pm0.14$ & 0.77 & $13.99\pm1.62$\\   \hline
2019& $F_{5100}$       & $6.19\pm0.23$ & 1.11 & $17.69\pm1.45$\\ 
    & $F_{\rm He~II}$  & $1.00\pm0.21$ & 0.74 & $70.77\pm6.05$\\ 
    & $F_{\rm He~I}$   & $1.98\pm0.09$ & 0.34 & $16.66\pm1.41$\\ 
    & $F_{\rm H\gamma}$& $2.51\pm0.14$ & 0.45 & $16.97\pm1.47$\\ 
    & $F_{\rm H\beta}$ & $7.16\pm0.18$ & 0.73 & $ 9.85\pm0.83$\\   \hline
2020& $F_{5100}$       & $7.90\pm0.29$ & 0.68 & $ 7.79\pm0.94$\\ 
    & $F_{\rm He~II}$  & $1.56\pm0.25$ & 0.38 & $19.01\pm3.14$\\ 
    & $F_{\rm He~I}$   & $2.32\pm0.11$ & 0.13 & $ 3.17\pm1.07$\\ 
    & $F_{\rm H\gamma}$& $3.00\pm0.16$ & 0.20 & $ 4.40\pm1.06$\\ 
    & $F_{\rm H\beta}$ & $8.67\pm0.21$ & 0.28 & $ 2.09\pm0.49$\\   \hline
2021& $F_{5100}$       & $6.84\pm0.39$ & 1.26 & $17.60\pm1.54$\\ 
    & $F_{\rm He~II}$  & $2.05\pm0.22$ & 1.14 & $55.09\pm4.52$\\ 
    & $F_{\rm He~I}$   & $2.34\pm0.13$ & 0.33 & $12.92\pm1.22$\\ 
    & $F_{\rm H\gamma}$& $3.46\pm0.19$ & 0.62 & $17.21\pm1.51$\\ 
    & $F_{\rm H\beta}$ & $8.58\pm0.33$ & 0.89 & $ 9.73\pm0.89$  
\enddata
\tablecomments{\footnotesize
The flux of $F_{\rm 5100}$ is in units of ${\rm 10^{-15}~erg~s^{-1}~cm^{-2}~\AA^{-1}}$, 
and the flux of broad emission line is in units of ${\rm 10^{-13}~erg~s^{-1}~cm^{-2}}$. 
$\sigma_{\rm LC}$ is the standard deviation of the light curve. 
The contamination of the host galaxy is eliminated. 
}
\end{deluxetable}

\begin{deluxetable*}{cccccccccccc}
  \tablecolumns{12}
  \tabletypesize{\scriptsize}
  \setlength{\tabcolsep}{4pt}
  \tablewidth{4pt}
  \tablecaption{Time lags in rest frame\label{tab_lag}}
  \tablehead{
\colhead{}  &
\multicolumn{2}{c}{$F_{\rm He~II}~vs.~F_{\rm 5100}$}         &
\colhead{}  &
\multicolumn{2}{c}{$F_{\rm He~I}~vs.~F_{\rm 5100}$}          &
\colhead{}  &
\multicolumn{2}{c}{$F_{\rm H\gamma}~vs.~F_{\rm 5100}$}  &
\colhead{}  &
\multicolumn{2}{c}{$F_{\rm H\beta}~vs.~F_{\rm 5100}$}       \\ \cline{2-3} \cline{5-6} \cline{8-9} \cline{11-12}
\colhead{Season}                 &
\colhead{$\tau_{\rm He~II}$ (days)}                        &
\colhead{$r_{\rm max}$}       &
\colhead{}  &
\colhead{$\tau_{\rm He~I}$ (days)}                        &
\colhead{$r_{\rm max}$}       &
\colhead{}  &
\colhead{$\tau_{\rm H\gamma}$ (days)}                        &
\colhead{$r_{\rm max}$}       &
\colhead{}  &
\colhead{$\tau_{\rm H\beta}$ (days)}                        &
\colhead{$r_{\rm max}$}       \\
\colhead{(1)}  &
\colhead{(2)}  &
\colhead{(3)}  &
\colhead{}  &
\colhead{(4)}  &
\colhead{(5)}  &
\colhead{}  &
\colhead{(6)}  &
\colhead{(7)}  &
\colhead{}  &
\colhead{(8)}  &
\colhead{(9)}  
}
\startdata
2015 &$ 0.65_{-0.94}^{+1.08} $& 0.85 &&$ 4.37_{-1.37}^{+1.65} $& 0.80 &&$ 5.09_{-2.20}^{+2.34} $& 0.70 &&$  7.20_{-0.35}^{+1.33} $& 0.83 \\ 
2018 &$-0.31_{-2.09}^{+1.78} $& 0.92 &&$ 3.31_{-2.19}^{+1.47} $& 0.96 &&$ 4.28_{-4.88}^{+1.86} $& 0.86 &&$  7.01_{-3.36}^{+2.33} $& 0.94 \\
2019 &$ 0.77_{-1.13}^{+0.84} $& 0.94 &&$ 5.47_{-0.86}^{+2.47} $& 0.92 &&$ 4.47_{-1.25}^{+1.50} $& 0.90 &&$  8.89_{-1.05}^{+2.03} $& 0.94 \\
2020 &$ 1.05_{-1.04}^{+1.55} $& 0.63 &&  --                                  & 0.26  &&   --                                 & 0.11 &&$ 10.03_{-3.27}^{+3.28} $& 0.56 \\
2021 &$ 1.30_{-0.88}^{+1.11} $& 0.95 &&$ 5.50_{-3.28}^{+1.75} $& 0.88 &&$ 4.55_{-2.26}^{+1.19} $& 0.88 &&$  9.02_{-2.48}^{+1.90} $& 0.81
\enddata
\tablecomments{\footnotesize
Time lags between the variations of broad-line fluxes and AGN continuum strength at 5100~\AA\ for each season. $r_{\rm max}$ is the maximum correlation coefficient. 
}
\end{deluxetable*}

\subsection{Time Lag} \label{sec:lag}
For each season, the time lags of the broad H$\gamma$, He~{\sc ii}, H$\beta$, and He~{\sc i} lines with respect to the varying AGN continuum at 5100~\AA\ are calculated using interpolation cross-correlation function (ICCF; \citealt{Gaskell1986,Gaskell1987}). 
Following \cite{Peterson2004}, the centroid of the ICCF above a typical value of $0.8~r_{\rm max}$ is assigned as the time lag, 
where $r_{\rm max}$ is the maximum cross-correlation coefficient. 
Monte Carlo simulations of random subset sampling and flux randomization are employed to construct the cross-correlation centroid distribution (CCCD), 
from which the uncertainty of time lag is estimated using the 15.87\% and 84.13\% quantiles. 

For the seasons of 2015, 2018, 2019, 2020, and 2021, the results of cross-correlation analysis are displayed in Figures~\ref{fig_map2015}$-$\ref{fig_map2021}, respectively. 
For each season, (1) the autocorrelation function (ACF) of the AGN continuum light curve is shown in 
panel~({\it aa}); (2) the CCFs between the light curves of AGN continuum and broad lines (in red) and corresponding CCCDs (in blue) are shown in panels~({\it ba}$-${\it ea}, hereafter CCF panels); (3) the rest-frame time lags measured from the above processes ($\tau_{\rm He~II}$, $\tau_{\rm He~I}$, $\tau_{\rm H\gamma}$, and $\tau_{\rm H\beta}$) 
are marked by the vertical dotted lines in CCF panels, and listed in Table~\ref{tab_lag}; 
(4) the maximum cross-correlation coefficients are noted in the CCF panels and also listed in Table~\ref{tab_lag}. 
For the season of 2020, 
because the maximum cross-correlation coefficients between the light curves of broad He~{\sc i} (and H$\gamma$ lines) and AGN continuum far less than 0.5, 
the time lags are not significant and therefore not listed. 
The five-season RM gives $\tau_{\rm He~II}:\tau_{\rm He~I}:\tau_{\rm H\gamma}:\tau_{\rm H\beta}=0.69:4.66:4.60:8.43$, 
which suggests a stratified BLR. 

\subsection{Line width}\label{sec:width}
The broad-line width is characterized by either the full width at half maximum (FWHM) or 
the line dispersion ($\sigma_{\rm line}$). In RM campaigns, these two parameters 
are usually calculated from both the mean and rms spectrum. It should be noted that 
the broad Balmer lines of NGC~5548 often have a double-peaked feature 
(see \citealt{Li2016}), which also exists throughout our spectroscopic monitoring periods. 
In the case that the double peaks are distinct, 
the FWHM of the double-peaked profile should be measured 
using the method described in \cite{Peterson2004}. 
Specifically, a blue-side peak and a red-side peak are defined on the line profile respectively, and the FWHM is taken as the wavelength separation of the two peaks. 

In each season, we first calculate the mean and rms spectrum of each broad line 
using its net broad lines constructed in (Section~\ref{sec:lc}), 
and then measure the line widths for each broad line, 
including FWHM (mean), $\sigma_{\rm line}$ (mean), 
FWHM (rms), and $\sigma_{\rm line}$ (rms). 
The broad He~{\sc ii} line is excluded because it is too weak to measure 
its line width reliably. 
We perform Monte Carlo simulations of random subset selection to generate 200 mean 
and rms spectra for each broad line, from which we construct four line-width distributions. 
Finally, the standard deviations calculated from each distribution are adopted as 
the uncertainties of line widths, respectively. 

In practice, instrument broadening is coupled with varying atmospheric (seeing) broadening. In this work, the average broadening of the broad emission line is estimated by
comparing the FWHM of [O~{\sc iii}]~$\lambda$5007 (410~km~$\rm s^{-1}$) from the high-resolution spectrum (see \citealt{Whittle1992}) with 
the averaged FWHM (807~km~$\rm s^{-1}$) from our whole campaign. 
This yields instrumental broadening of 695~km~$\rm s^{-1}$ (in the FWHM case), 
corresponding to 295~km~$\rm s^{-1}$ (in $\sigma_{\rm line}$ case) for a Gaussian model of the [O~{\sc iii}] line profile. 
The broadening-corrected line widths are listed in Table~\ref{tab_lw}, 
where the broad H$\beta$ line widths from the season of 2015 reported in \cite{Lu2016} are updated. 
These measurements are used to investigate the virial relation of the BLR in Section~\ref{sec:vr}. 

\begin{deluxetable*}{cccccccc}
  \tablecolumns{8}
  \tabletypesize{\scriptsize}
  \setlength{\tabcolsep}{4pt}
  \tablewidth{4pt}
  \tablecaption{Line width\label{tab_lw}}
  \tablehead{
\colhead{}   &
\multicolumn{3}{c}{FWHM (${\rm km~s^{-1}}$)}         &
\colhead{}  &
\multicolumn{3}{c}{$\sigma_{\rm line}$ (${\rm km~s^{-1}}$)}         \\ \cline{2-4} \cline{6-8}  
\colhead{Season}   &
\colhead{H$\gamma$}  &
\colhead{H$\beta$}  &
\colhead{He~I}  &
\colhead{}  &
\colhead{H$\gamma$}  &
\colhead{H$\beta$}  &
\colhead{He~I}  \\
\colhead{(1)}   &
\colhead{(2)}  &
\colhead{(3)}  &
\colhead{(4)}  &
\colhead{}  &
\colhead{(5)}  &
\colhead{(6)}  &
\colhead{(7)}  
}
\startdata
\multicolumn{8}{c}{Mean spectrum}\\ \cline{1-8}
2015& $ 11671 \pm  788 $&$ 11623 \pm  352 $&$  8898 \pm  393 $&&$  3781 \pm   60 $&$  4307 \pm  150 $&$  3546 \pm  107 $\\
2018& $ 11415 \pm 1438 $&$ 12221 \pm  535 $&$ 11231 \pm 1288 $&&$  3266 \pm   16 $&$  3881 \pm  111 $&$  3960 \pm  110 $\\
2019& $ 11171 \pm 1229 $&$ 10493 \pm  258 $&$ 10635 \pm   51 $&&$  3338 \pm  217 $&$  3717 \pm   94 $&$  3770 \pm  212 $\\
2020& $ 10343 \pm 1494 $&$  9657 \pm  617 $&$ 10335 \pm  270 $&&$  3171 \pm  172 $&$  3758 \pm   14 $&$  3783 \pm  260 $\\
2021& $ 10348 \pm 1051 $&$  9578 \pm  159 $&$  9977 \pm  342 $&&$  3097 \pm  135 $&$  3574 \pm   28 $&$  3664 \pm  202 $\\
\hline
\multicolumn{8}{c}{rms spectrum}\\ \cline{1-8}
2015& $ 11258 \pm  679 $&$ 10241 \pm  515 $&$ 12180 \pm  789 $&&$  4063 \pm  387 $&$  4377 \pm  477 $&$  3650 \pm  278 $\\
2018& $  9912 \pm  291 $&$  9724 \pm  599 $&$ 12576 \pm  499 $&&$  4214 \pm  450 $&$  3989 \pm  429 $&$  4077 \pm  123 $\\
2019& $  9919 \pm  820 $&$  9053 \pm  710 $&$  8223 \pm 1915 $&&$  3730 \pm  441 $&$  3732 \pm  300 $&$  3674 \pm  116 $\\
2020& $  9534 \pm  528 $&$  9199 \pm  863 $&$ 11025 \pm  398 $&&$  3719 \pm  455 $&$  3209 \pm  600 $&$  3728 \pm  333 $\\
2021& $  9710 \pm 1124 $&$  8470 \pm  423 $&$ 10125 \pm  611 $&&$  3757 \pm  302 $&$  3349 \pm  436 $&$  3285 \pm  126 $
\enddata
\tablecomments{\footnotesize
The broad-line widths including FWHM and $\sigma_{\rm line}$ are measured from mean and rms spectra for each season. 
The contamination of other blended components is eliminated by spectral fitting and decomposition. 
}
\end{deluxetable*}

\subsection{Velocity-resolved Reverberation Mapping}\label{sec:vr}
Following the previous works (e.g., \citealt{Denney2010,Bentz2009,Grier2013,Du2016,Lu2016,Pei2017}), 
we carry out velocity-resolved RM analysis using the net broad H$\gamma$ 
and H$\beta$ lines obtained in Section~\ref{sec:lc}. For each season, 
we first calculate the rms spectrum of the net broad lines 
and select an emission-line window from the rms spectrum. 
Then we divide the broad lines into several uniformly spaced bins within the selected window
and measure the velocity-binned light curves by integrating the fluxes in each velocity bin. 
Finally, we calculate the time lags between the velocity-binned light curves 
and the AGN continuum at 5100~\AA~using the same procedures as in Section~\ref{sec:lag}. 

\begin{figure*}[htp]
\includegraphics[width=0.93\textwidth]{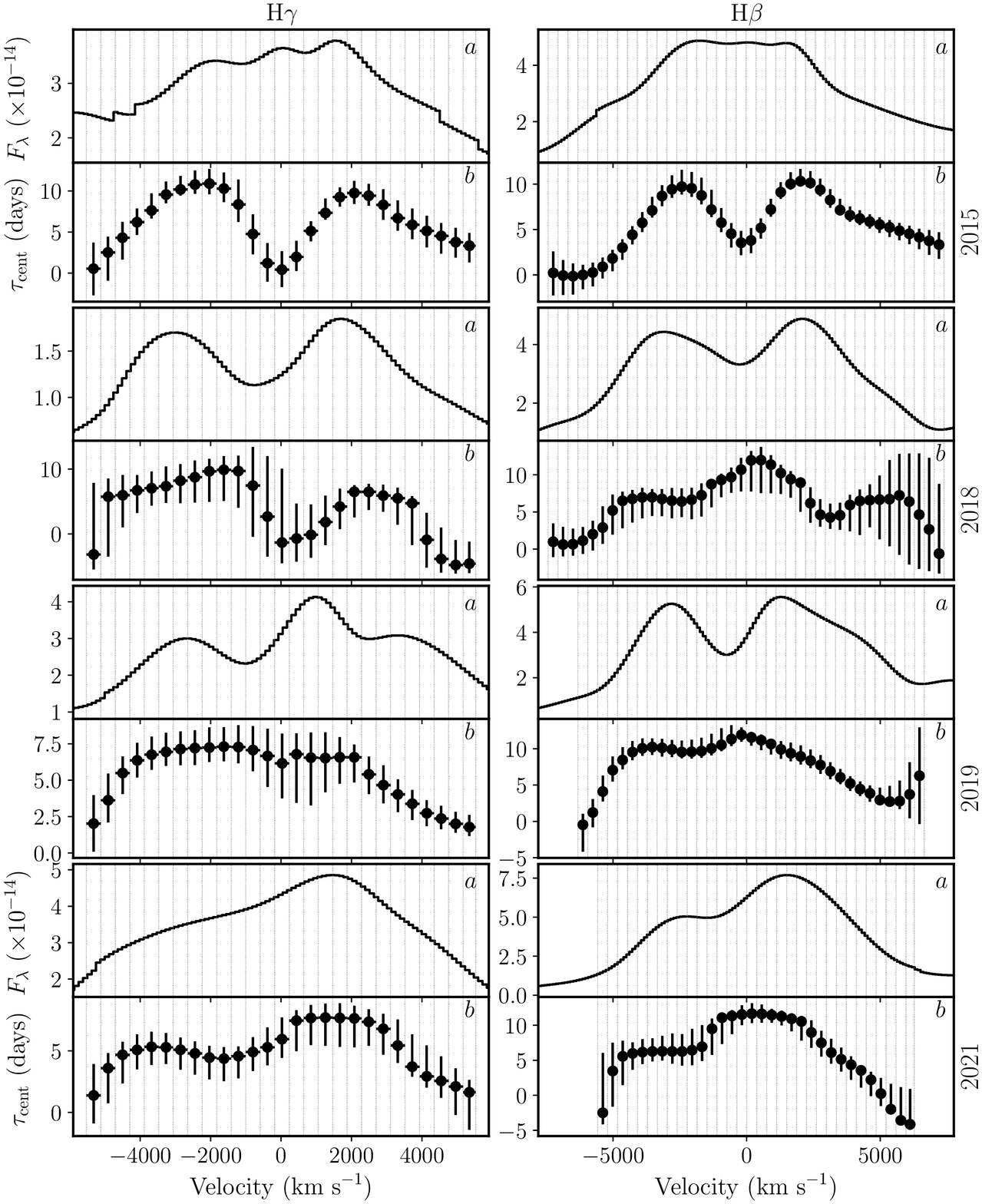}
\caption{
The results of velocity-resolved RM analysis for the seasons of 2015, 2018, 2019, and 2021. 
For each season, 
panel~({\it a}) shows the rms spectrum of broad H$\gamma$ (left) or H$\beta$ (right), 
and panel~({\it b}) shows the centroid time lags as a function of velocity in the line of sight. 
The vertical dotted lines are the edges of the velocity bins. 
The H$\gamma$ and H$\beta$ velocity-resolved lag profiles are plotted in the same coordinate 
ranges in the direction of the velocity axis, respectively. 
All maximum cross-correlation coefficients between the velocity-binned light curves and the varying 
AGN continuum at 5100~\AA\ are larger than 0.5. 
Note that the broad-line velocity binning is below the instrumental resolution of $\sim$695~km~s$^{-1}$, and thus the measurements are not independent. 
}
\label{fig:ljmaps}
\end{figure*}

For the seasons of 2015, 2018, 2019, and 2021, 
the results of velocity-resolved RM analysis are displayed in Figure~\ref{fig:ljmaps}. 
The top panels ({\it a}) show the rms spectra of the net broad H$\gamma$ (left) and 
H$\beta$ (right) lines. 
The bottom panels ({\it b}) show the broad H$\beta$ and H$\gamma$ velocity-resolved lag profiles 
(VRLP; i.e., velocity-resolved time lags as a function of line-of-sight velocity, 
hereafter H$\beta$ VRLP and H$\gamma$ VRLP). 
The above velocity-binned light curves are measured from the fitted broad lines. 
Alternatively, the H$\beta$ and H$\gamma$ VRLPs can also be measured from the net broad lines 
constructed by subtracting other blended components from the calibrated spectra. 
We find that the results from the above approaches are in agreement. 
For the season of 2020, we do not obtain credible H$\beta$ and H$\gamma$ VRLPs 
due to the very low variability (see Figure~\ref{fig_map2020}). 
The broad Helium lines are not considered for the velocity-resolved RM analysis because of very weak fluxes. 

Many theoretical models studied the geometry and kinematics of BLR 
(e.g., \citealt{Horne2004,Grier2013,Gaskell1988,Gaskell2013}). 
A basic notion is that an outflowing BLR would lead to a longer time lag in the red wing 
than in the blue wing of the line profile (i.e., blue-leads-red). 
For an inflowing BLR, the situation is just opposite (i.e., red-leads-blue, e.g., see Figure 14 of \citealt{Grier2013}). 
For BLR gas with Keplerian motion \citep{Welsh1991}, the shortest lag would be in the line wings, 
and the longest lag would be in the line core if the ionizing source 
is emitting isotropically, because the gas with a higher velocity is closer to the 
central SMBH. If the BLR gas is illuminated anisotropically by a central ionizing source, 
a flat disc or spherical geometry structure of the BLR in Keplerian orbits could give rise 
to a double-peaked VRLP \citep{Welsh1991,Goad1996}. 
Additionally, many works based on different methods have pointed to the scenario 
that the BLR has disc-shaped geometry \citep{Goad1996,Wills1995,McLure2002,Down2010}. 
Using equivalent width of the [O~{\sc iii}]~as indicator of accretion disc inclination, 
\cite{Bisogni2017b} further confirmed this scenario (see also \citealt{Bisogni2017a,Risaliti2011}). 
Keep these theoretical models of BLR kinematics in mind, 
below we give a brief description of VRLPs. 

In the season of 2015, both the H$\gamma$ and H$\beta$ VRLPs present 
an `M'-shaped (or a double-peaked) structure (see Figure~\ref{fig:ljmaps}). 
This structure was also observed in 2014 (see H$\beta$ VRLP of \citealt{Pei2017}) 
and its possible origin had been discussed by \cite{Horne2021}. 
According to the theoretical model of the BLR kinematics \citep{Welsh1991,Goad1996} 
and the suggestion about the BLR geometry \citep{McLure2002,Bisogni2017b,Risaliti2011}, 
we propose that the BLR of NGC~5548 is a virialized flat disc and illuminated 
anisotropically by the central ionizing source during the observation periods. 
In our previous work of \cite{Lu2016}, a simple H$\beta$ VRLP was obtained by dividing 
the rms spectrum of the broad H$\beta$ line into nine uniformly spaced bins, 
which presents a nearly symmetric structure, but the lag of the line-core bin is shorter than adjacent bins. 
This is actually consistent with the reconstructed H$\beta$ VRLP in this work. 
In the season of 2018, the H$\gamma$ VRLP presents a distorted `M'-shaped structure whereas the `M'-shaped structure of H$\beta$ VRLP disappears. 
In the seasons of 2019 and 2021, the H$\gamma$ and H$\beta$ VRLPs generally present 
the red-leads-blue trend, but the VRLPs in the red and blue wings seemingly 
have the same time lags. These complex signatures could be caused by a mix of infalling and 
virialized motions according to the above theoretical models, 
or caused by a switch between the infalling and virialized as proposed by \cite{Xiao2018}. 
Alternatively, the accretion disc winds/outflows with different transparency could 
partially shield or obscure the central ionizing source 
(e.g, \citealt{Mangham2017,Mangham2019,Gaskell2018,Dehghanian2019}), 
leading to anisotropic illuminations onto the BLR and hence giving rise to complex signatures of VRLPs. 

In addition, by inspecting Figure~\ref{fig:ljmaps}, we find that the VRLPs are changing with the seasons.  
More interestingly, by comparing the VRLPs from the seasons of 2015 and 2018, 
we find that the changes of H$\gamma$ VRLPs lag behind the H$\beta$ VRLPs, 
the reason is that the `M'-shaped structure of H$\gamma$ and H$\beta$ VRLPs presented in 2015 
changed as a distorted `M'-shaped structure in the H$\gamma$ VRLPs and disappeared in the H$\beta$ VRLPs in 2018. 
These findings perhaps indicate a change in the kinematics of BLR. 
Using the dynamical modeling approach developed by \cite{Pancoast2014} to further constrain 
the geometry and kinematics of BLR (see also \citealt{Pancoast2011,Li2013,Li2018}), 
will be presented in future contributions. 

\section{Properties of the BLR in NGC~5548}\label{sec:prop}
Since 1988, there were 18 seasons of RM measurements for NGC~5548 from different spectroscopic monitoring campaigns. 
These campaigns include the AGN Watch project (from 1988 to 2001, see \citealt{Peterson1993,Peterson1998,Peterson1999,Peterson2002}), 
the LAMP AGN project (The Lick AGN Monitoring Project, in 2008, see \citealt{Bentz2009}), 
the AGN STROM project (Space Telescope and Optical Reverberation Mapping Project, in 2014, 
see \citealt{Pei2017}), and three individual RM campaigns which were undertaken mainly by the 
McGraw-Hill 1.3 m telescope at the MDM Observatory (in 2005, 2007, and 2012; 
see \citealt{Bentz2007}, \citealt{Denney2010}, and \citealt{DeRosa2018}). 
Our five seasons of RM measurements are based on the long-term spectroscopic monitoring project with the Lijiang 2.4~m telescope. 
As a result, thus far there are in total 23 seasons of RM measurements for NGC~5548, making it to be the most intensively RM monitored AGN.

In Table~\ref{tab_sum}, we compile the previous RM measurements of NGC~5548 
along with our five-season RM results. The annual averaged AGN continuum flux 
$F_{\rm 5100}$ listed in Table~\ref{tab_sum} had been corrected for the contamination 
of the host galaxy, where the values of the first 16 rows are from \cite{Eser2015}, 
the 17th row is compiled from \cite{DeRosa2018} and 
updated by eliminating the host-galaxy flux, 
the 18th row is from \cite{Pei2017}, and the last 5 rows are from this work. 
We calculate the standard deviations ($\delta_{F_{\rm 5100}}$) of the AGN continuum light curves for each campaign and list them in Table~\ref{tab_sum}, 
which are used to estimate the error of the optical luminosity at 5100~\AA\ in Section~\ref{sec:rl}. 
We also list the line dispersion of $\sigma_{\rm line}$ measured from the rms spectrum 
and the time lag of the broad H$\beta$ line. These quantities are used to calculate the virial SMBH mass in Section~\ref{sec:mass}. 
Next, based on 23 RM measurements of NGC 5548, 
we construct the virial relation and the BLR radius$-$luminosity relation of NGC~5548, 
and investigate the stability of the BLR. 

\begin{deluxetable*}{cccccccccc}
  \tablecolumns{10}
  \tabletypesize{\scriptsize}
  \setlength{\tabcolsep}{4pt}
  \tablewidth{4pt}
  \tablecaption{Summary of RM measurements for NGC~5548\label{tab_sum}}
  \tablehead{
\colhead{Data set}    &
\colhead{Dates}    &
\colhead{$F_{\rm 5100}$}    &
\colhead{$\delta_{F_{\rm 5100}}$}    &
\colhead{$\tau_{\rm H\beta}$}    &
\colhead{$\sigma_{\rm line}$~(rms)}    &
\colhead{Virial~Product}    &
\colhead{$M_{\rm BH}$}    &
\colhead{$\dot{\mathscr{M}}$}    &
\colhead{References}  \\
\colhead{}    &
\colhead{}    &
\colhead{($\times10^{-15}$)}    &
\colhead{($\times10^{-15}$)}    &
\colhead{(days)}    &
\colhead{(km~s$^{-1}$)}    &
\colhead{($\times10^{7}M_{\odot}$)}    &
\colhead{($\times10^{7}M_{\odot}$)}    &
\colhead{}    &
\colhead{}  \\ 
\colhead{(1)}    &
\colhead{(2)}    &
\colhead{(3)}    &
\colhead{(4)}    &
\colhead{(5)}    &
\colhead{(6)}    &
\colhead{(7)}    &
\colhead{(8)}    &
\colhead{(9)}    &
\colhead{(10)}  
}
\startdata
\multicolumn{10}{c}{Previous campaign}\\ \cline{1-10}
Year  1 & 1988 Dec$-$1989 Oct & $6.18\pm0.65$ & 1.26 & $19.70_{-1.50}^{+1.50}$ & $1687\pm 56$ & $1.10_{-0.11}^{+0.11}$ & $ 6.90_{-0.70}^{+0.70}$ & 0.059 & (1, 2, 3)\\
Year  2 & 1989 Dec$-$1990 Oct & $3.38\pm0.55$ & 1.00 & $18.60_{-2.30}^{+2.10}$ & $1882\pm 83$ & $1.29_{-0.20}^{+0.18}$ & $ 8.11_{-1.23}^{+1.16}$ & 0.017 & (1, 2, 3)\\
Year  3 & 1990 Nov$-$1991 Oct & $5.34\pm0.55$ & 0.92 & $15.90_{-2.50}^{+2.90}$ & $2075\pm 81$ & $1.34_{-0.23}^{+0.27}$ & $ 8.42_{-1.48}^{+1.67}$ & 0.032 & (1, 2, 3)\\
Year  4 & 1992 Jan$-$1992 Oct & $2.90\pm0.50$ & 1.16 & $11.00_{-2.00}^{+1.90}$ & $2264\pm 88$ & $1.10_{-0.22}^{+0.21}$ & $ 6.94_{-1.37}^{+1.31}$ & 0.019 & (1, 2, 3)\\
Year  5 & 1992 Nov$-$1993 Sep & $5.38\pm0.55$ & 0.86 & $13.00_{-1.40}^{+1.60}$ & $1909\pm129$ & $0.93_{-0.16}^{+0.17}$ & $ 5.83_{-1.01}^{+1.07}$ & 0.067 & (1, 2, 3)\\
Year  6 & 1993 Nov$-$1994 Oct & $5.62\pm0.59$ & 1.10 & $13.40_{-4.30}^{+3.80}$ & $2895\pm114$ & $2.19_{-0.72}^{+0.65}$ & $13.82_{-4.57}^{+4.07}$ & 0.013 & (1, 2, 3)\\
Year  7 & 1994 Nov$-$1995 Oct & $7.92\pm0.51$ & 1.00 & $21.70_{-2.60}^{+2.60}$ & $2247\pm134$ & $2.14_{-0.36}^{+0.36}$ & $13.48_{-2.28}^{+2.28}$ & 0.022 & (1, 2, 3)\\
Year  8 & 1995 Nov$-$1996 Oct & $6.02\pm0.55$ & 1.64 & $16.40_{-1.10}^{+1.20}$ & $2026\pm 68$ & $1.31_{-0.12}^{+0.13}$ & $ 8.28_{-0.79}^{+0.82}$ & 0.039 & (1, 2, 3)\\
Year  9 & 1996 Dec$-$1997 Oct & $3.77\pm0.51$ & 0.91 & $17.50_{-1.60}^{+2.00}$ & $1923\pm 62$ & $1.26_{-0.14}^{+0.17}$ & $ 7.96_{-0.89}^{+1.04}$ & 0.021 & (1, 2, 3)\\
Year 10 & 1997 Nov$-$1998 Sep & $8.34\pm0.63$ & 1.45 & $26.50_{-2.20}^{+4.30}$ & $1732\pm 76$ & $1.55_{-0.19}^{+0.29}$ & $ 9.78_{-1.18}^{+1.80}$ & 0.046 & (1, 2, 3)\\
Year 11 & 1998 Nov$-$1999 Oct & $6.90\pm0.60$ & 1.82 & $24.80_{-3.00}^{+3.20}$ & $1980\pm 30$ & $1.90_{-0.24}^{+0.25}$ & $11.96_{-1.49}^{+1.59}$ & 0.023 & (1, 2, 3)\\
Year 12 & 1999 Dec$-$2000 Sep & $2.41\pm0.50$ & 1.20 & $ 6.50_{-3.70}^{+5.70}$ & $1969\pm 48$ & $0.49_{-0.28}^{+0.43}$ & $ 3.10_{-1.77}^{+2.72}$ & 0.071 & (1, 2, 3)\\
Year 13 & 2000 Nov$-$2001 Dec & $2.32\pm0.51$ & 0.86 & $14.30_{-4.30}^{+5.90}$ & $2173\pm 89$ & $1.32_{-0.41}^{+0.55}$ & $ 8.31_{-2.59}^{+3.49}$ & 0.009 & (1, 2, 3)\\
Year 17 & 2005 Mar$-$2005 Apr & $0.97\pm0.53$ & 0.53 & $ 6.30_{-2.30}^{+2.60}$ & $2939\pm373$ & $1.06_{-0.47}^{+0.51}$ & $ 6.70_{-2.98}^{+3.24}$ & 0.004 & (3, 4)\\
Year 19 & 2007 Mar$-$2007 Jul & $1.35\pm0.48$ & 0.48 & $ 5.07_{-2.46}^{+2.37}$ & $1822\pm 35$ & $0.33_{-0.16}^{+0.15}$ & $ 2.07_{-1.01}^{+0.97}$ & 0.067 & (3, 5)\\
Year 20 & 2008 Feb$-$2008 Jun & $1.21\pm0.41$ & 0.10 & $ 4.17_{-1.33}^{+0.90}$ & $4270\pm292$ & $1.48_{-0.52}^{+0.38}$ & $ 9.36_{-3.25}^{+2.39}$ & 0.003 & (3, 6)\\
Year 23 & 2012 Jan$-$2012 Apr & $3.44\pm0.12$ & 0.56 & $ 2.83_{-0.96}^{+0.88}$ & $2772\pm 33$ & $0.42_{-0.14}^{+0.13}$ & $ 2.68_{-0.91}^{+0.83}$ & 0.163 & (7)\\
Year 25 & 2014 Jan$-$2014 Jul & $7.44\pm0.50$ & 0.80 & $ 4.17_{-0.36}^{+0.36}$ & $4278\pm671$ & $1.49_{-0.48}^{+0.48}$ & $ 9.39_{-3.06}^{+3.06}$ & 0.042 & (8)\\
\hline
\multicolumn{10}{c}{Our five-season campaign}\\ \cline{1-10}
Year 26 & 2015 Jan$-$2015 Aug & $4.34\pm0.30$ & 1.05 & $ 7.20_{-0.35}^{+1.33}$ & $4377\pm477$ & $2.69_{-0.60}^{+0.77}$ & $16.97_{-3.79}^{+4.85}$ & 0.006 & (9, 10)\\
Year 29 & 2018 Mar$-$2018 Jun & $5.13\pm0.20$ & 0.77 & $ 7.01_{-3.36}^{+2.33}$ & $3989\pm429$ & $2.18_{-1.14}^{+0.86}$ & $13.72_{-7.21}^{+5.43}$ & 0.011 & (10)\\
Year 30 & 2018 Nov$-$2019 Jun & $6.19\pm0.23$ & 1.11 & $ 8.89_{-1.05}^{+2.03}$ & $3732\pm300$ & $2.42_{-0.48}^{+0.68}$ & $15.24_{-3.04}^{+4.25}$ & 0.012 & (10)\\
Year 31 & 2020 Jan$-$2020 Jun & $7.90\pm0.29$ & 0.68 & $10.03_{-3.27}^{+3.28}$ & $3209\pm600$ & $2.02_{-1.00}^{+1.00}$ & $12.71_{-6.30}^{+6.31}$ & 0.025 & (10)\\
Year 32 & 2020 Dec$-$2021 Aug & $6.84\pm0.39$ & 1.26 & $ 9.02_{-2.48}^{+1.90}$ & $3349\pm436$ & $1.98_{-0.75}^{+0.66}$ & $12.45_{-4.71}^{+4.17}$ & 0.021 & (10)
\enddata
\tablecomments{\footnotesize
The AGN continuum flux at 5100~\AA\ ($F_{\rm 5100}$) and its standard deviation 
($\delta_{F_{\rm 5100}}$) are in units of ${\rm erg~s^{-1}~cm^{-2}~\AA^{-1}}$, 
which are used to calculate the optical luminosity at 5100~\AA\ and its uncertainty. 
The line dispersion ($\sigma_{\rm line}$) measured from the rms spectrum of broad H$\beta$ line and the H$\beta$ time lag are used to 
calculate the virial product and virial SMBH mass ($M_{\rm BH}$). 
Other line widths including FWHM and $\sigma_{\rm line}$ measured from mean and rms spectrum are referred to Table~4 of \cite{Lu2016}, \cite{DeRosa2018} and \cite{Pei2017}. 
The dimensionless accretion rate $\dot{\mathscr{M}}$ is estimated using the optical luminosity and SMBH mass. \\
{\bf References.} (1) \cite{Eser2015}, (2) \cite{Collin2006}, (3) \cite{Bentz2013}, (4) \cite{Bentz2007}, (5) \cite{Denney2010}, (6) \cite{Bentz2009}, (7) \cite{DeRosa2018}, (8) \cite{Pei2017}, 
(9) \cite{Lu2016}, (10) This work
}
\end{deluxetable*}

\subsection{The Virial Relation}\label{sec:vr}
\begin{figure*}[htb]
\centering
\includegraphics[angle=0,width=0.93\textwidth]{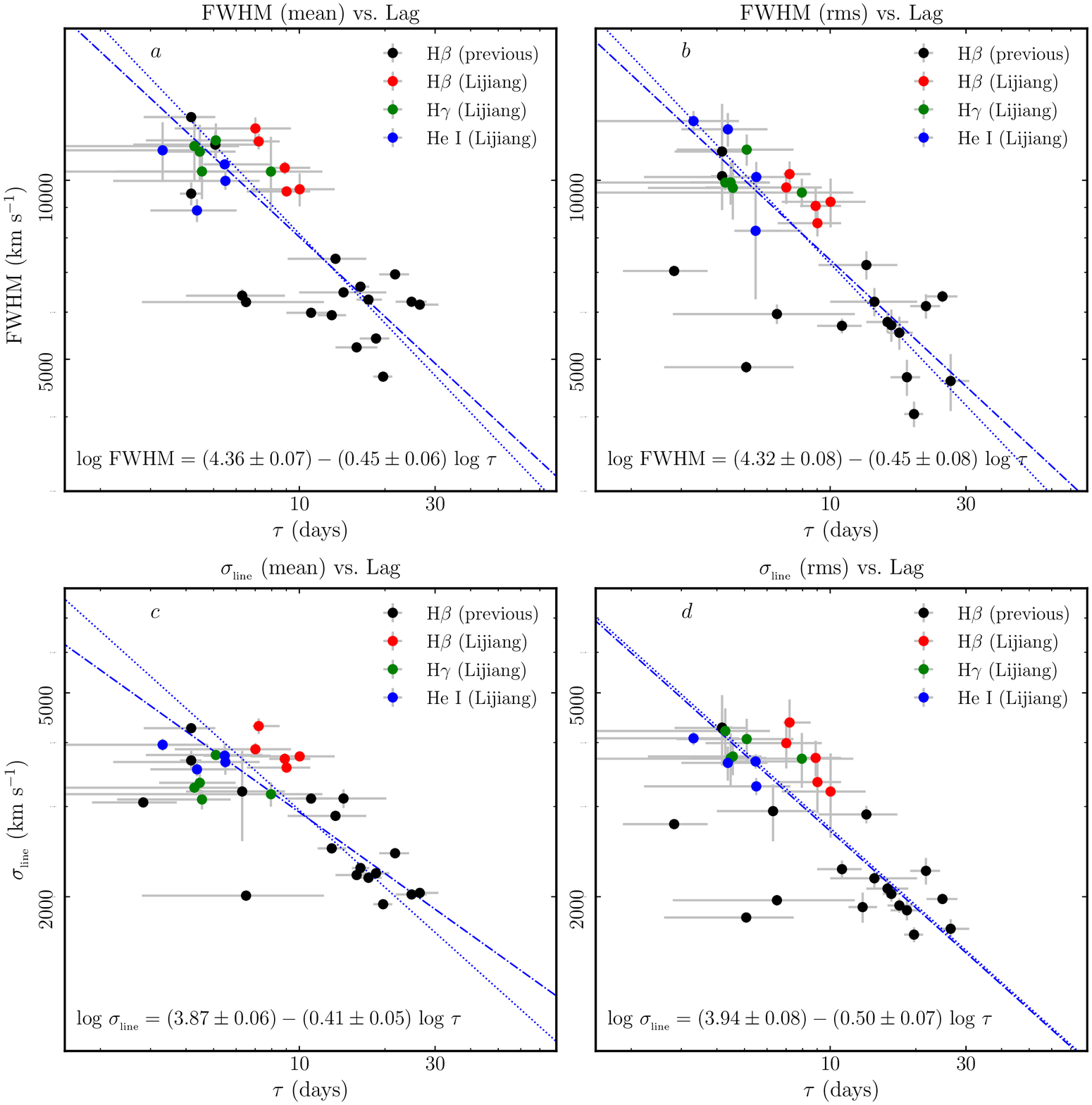}
\caption{\footnotesize
Relation between the broad-line widths and their time lags for NGC~5548, 
the colorful circles are the new measurements from this work, and the black circles are the previous measurements. 
The broad-line width can be characterized by either the FWHM or the line dispersion ($\sigma_{\rm line}$), 
and both can be measured from the mean or rms spectra. For each case, 
the dot-dashed lines are the fit to the relation of ${\rm log}~({\rm FWHM~or}~\sigma_{\rm line})=a+b~{\rm log}~\tau$, 
the results of best fit are recorded at the bottom of each panel. 
The dotted lines are the best fit with a forced virial slope of $b=-0.5$. 
}
\label{fig_vr}
\end{figure*}

The BLR lies within the sphere of influence of the central SMBH, therefore, its kinematics is expected to be dominated by the gravitational potential of the SMBH. 
This physical property is usually tested through the virial relation, 
that is $V\propto\tau^{-0.5}$ relation, 
where $V$ and $\tau$ are the line width and time lag of a broad emission line, respectively. 
For example, the virial relations of the BLR in Mrk~817 and NGC~7469 were investigated by \cite{Lu2021a} using the archival data. 
In NGC~5548, the virial relation of the BLR was successively investigated using multi-season RM results by \cite{Peterson2004}, \cite{Bentz2007}, and \cite{Lu2016}. 
By adding our five-season RM, we update the virial relation in Figure~\ref{fig_vr}. 
It is possible that the line width measured from the rms spectrum (i.e., variable spectrum) just represents the projected velocity of 
the BLR that with broad-line variability (i.e., the variable region), 
while the line width measured from the mean spectrum actually represents the projected 
velocity of the whole BLR that with the contributions of broad-line flux. 
Therefore, both FWHM and $\sigma_{\rm line}$ measured from mean and rms spectra, 
that is FWHM (mean), FWHM (rms), $\sigma_{\rm line}$ (mean), and $\sigma_{\rm line}$ (rms), 
are considered for comparison. 

In Figure~\ref{fig_vr}, the dot-dashed line is the best fit to the relation of ${\rm log}~({\rm FWHM~or}~\sigma_{\rm line})=a+b~{\rm log}~\tau$. 
The yielded slope $b=-0.45\pm0.06$ for FWHM (mean) $vs.$ time lag, $b=-0.45\pm0.08$ for FWHM (rms) $vs.$ time lag, 
$b=-0.41\pm0.05$ for $\sigma_{\rm line}$ (mean) $vs.$ time lag, and $b=-0.50\pm0.07$ for $\sigma_{\rm line}$ (rms) $vs.$ time lag, 
all with a mean intrinsic scatter of 0.26~dex. 
These best-fit results are also marked in the corresponding panel of Figure~\ref{fig_vr}. 
The measurements from our five-season RM adequately expand the dynamical range of parameters. 
The dotted lines are the fit with a theoretical virial slope of $b=-0.5$ for each case. 
All slopes obtained from the above virial relations are close to the theoretical virial slope, 
generally confirming that the whole BLR is bounded by the gravitational potential of the central SMBH. 

\subsection{The Radius$-$Luminosity Relation}\label{sec:rl}
\begin{figure}[htb]
\centering
\includegraphics[angle=0,width=0.48\textwidth]{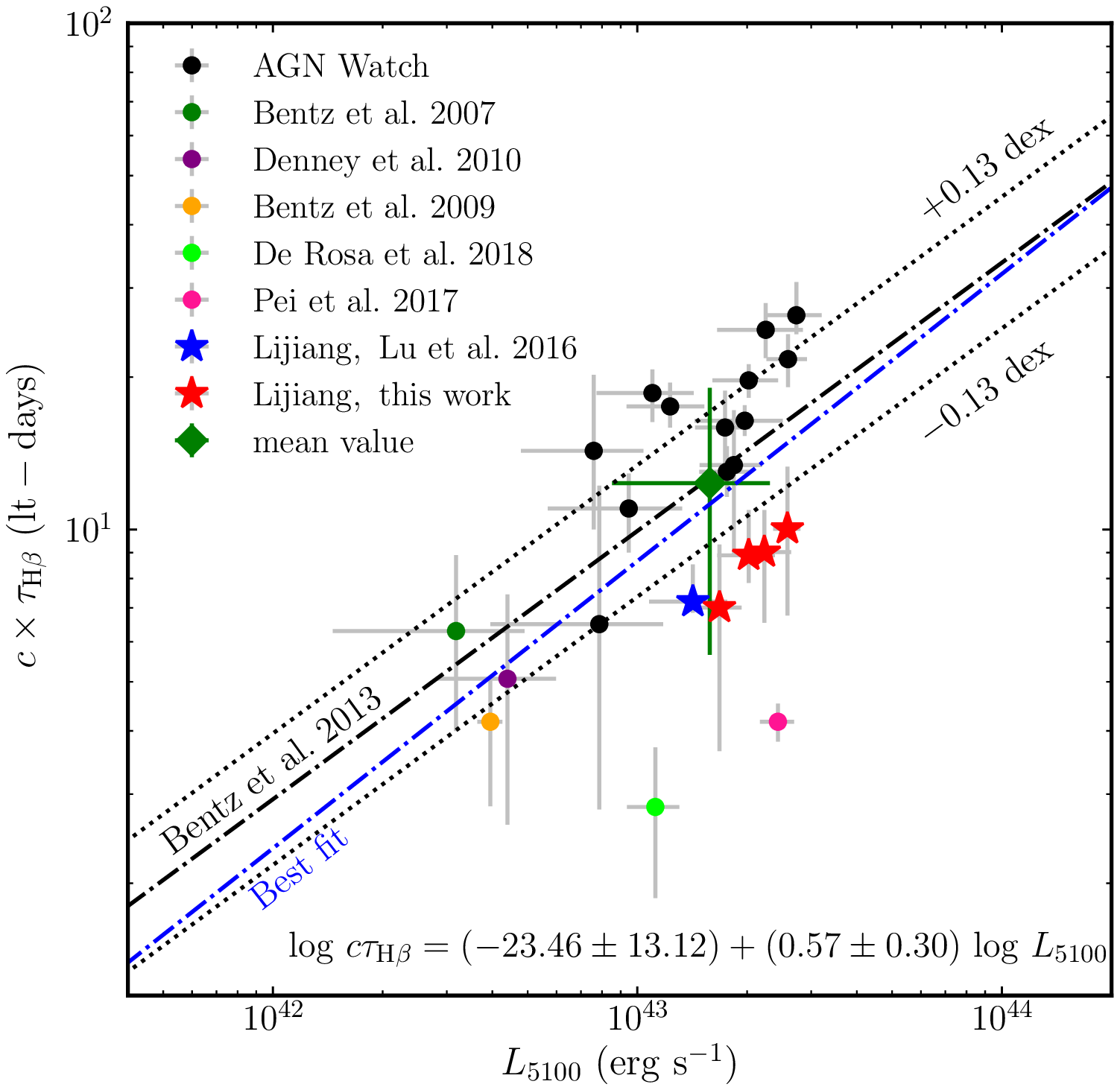}
\caption{\footnotesize
Relation between the optical luminosity at 5100~\AA~and the H$\beta$ time lag 
from 23 RM campaigns for NGC~5548 (see the inserted legend). 
The black dot-dashed line is the canonical $R_{\rm H\beta}-L_{\rm 5100}$ relation 
with a slope of 0.53 (\citealt{Bentz2013}), 
and the dotted lines are the intrinsic scatter ($\pm$0.13~dex) of this relation. 
The blue dot-dashed line is the best fit for data of NGC~5548, 
which gives the slope of 0.57 with an intrinsic scatter of 0.26~dex, 
the fitted relation is recorded at the bottom of the figure. 
The averaged values of $\tau_{\rm H\beta}$ and $L_{\rm 5100}$ over 23 campaigns (green diamond) are 
consistent with the canonical $R_{\rm H\beta}-L_{\rm 5100}$ relation. 
The previous RM measurements are shown in the symbol of circles, and our five-season RM measurements are shown in ``$\bigstar$". 
The different symbols with different colors are used to distinguish the different campaigns. 
}
\label{fig_rl}
\end{figure}

In Figure~\ref{fig_rl}, we investigate the BLR radius$-$luminosity 
($R_{\rm H\beta}-L_{\rm 5100}$) relation of NGC~5548. 
$R_{\rm H\beta}=c\tau_{\rm H\beta}$ is the BLR radius (or size), where $c$ is the speed of light and 
$\tau_{\rm {H\beta}}$ is the time lag of broad H$\beta$ line. 
The optical luminosity $L_{\rm 5100}$ is derived from the AGN continuum flux 
at 5100~\AA\ ($F_{\rm 5100}$) and listed in Table~\ref{tab_sum}, 
As mentioned above, the error of $L_{\rm 5100}$ is estimated from 
the standard deviation of the $F_{\rm 5100}$ (i.e., $\delta_{F_{\rm 5100}}$). 
We fit the results with the relation of 
${\rm log}~c\tau_{\rm H\beta}=a+b~{\rm log}~L_{\rm 5100}$, 
yielding a slope of $b=0.57\pm0.30$. 
The best-fit relation is displayed with the blue dot-dashed line in Figure~\ref{fig_rl}. 
By adding our five-season RM measurements, this slope is far less than the value of 0.86 reported 
by \cite{Lu2016} and almost consistent with the slope (0.53) of canonical 
$R_{\rm H\beta}-L_{\rm 5100}$ relation (see \citealt{Bentz2013}). 

For comparison, the canonical $R_{\rm H\beta}-L_{\rm 5100}$ relation with a slope of 0.53 and intrinsic scatter of 0.13~dex (see \citealt{Bentz2013}) are also plotted in Figure~\ref{fig_rl}. 
We find that although our RM measurements overall lie $0.13$~dex below the canonical relation, 
the slope lines up with that of the canonical relation. 
From the 23 RM campaigns of NGC~5548, 
we calculate the mean H$\beta$ time lag of 12.3~days with a standard deviation of 6.7~days, 
and mean optical luminosity of $1.58\times10^{43}~{\rm erg~s^{-1}}$ with a standard deviation of $0.73\times10^{43}~{\rm erg~s^{-1}}$. 
By superimposing the mean values of time lag and optical luminosity in Figure~\ref{fig_rl}, we find that the location is well consistent with the 
canonical $R_{\rm H\beta}-L_{\rm 5100}$ relation. 

\begin{figure*}[ht!]
\centering
\includegraphics[angle=0,width=0.98\textwidth]{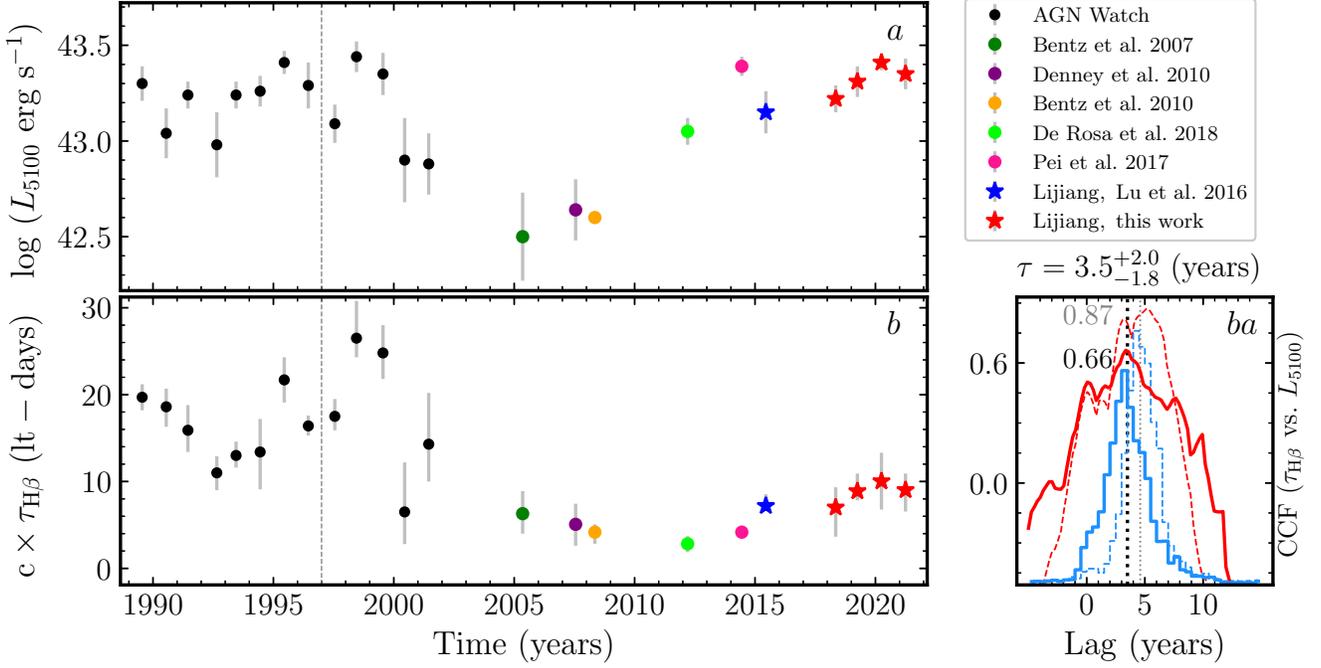}  %{5100vstotal_lag_nn.eps}  {5100vstotal_lagsigma.eps} {Lag_year2.eps}
\caption{\footnotesize
Variations of the optical luminosity (panel {\it a}) and the BLR radius (panel {\it b}) for NGC~5548 measured between 1988 and 2021, 
and the result of cross-correlation analysis (panel {\it ba}). 
The different symbols with different colors are used in the left panels to distinguish the different campaigns. 
In panel ({\it ba}), the red solid curve is the CCF between the BLR radius and the optical luminosity measured between 1988 and 2021 ($r_{\rm max}=0.66$), 
and the blue solid histogram is the cross-correlation centroid distribution (CCCD), the measured time lag of $3.5^{+2.0}_{-1.8}$~years is marked by the vertical dotted line in black. 
A sub-set of the BLR radius and the optical luminosity measured after 1997 (marked as the vertical dashed line in panel~{\it a} and {\it b}) 
is selected to recalculate the time lag. The result is also shown in panel~({\it ba}), 
where the red dashed curve is the CCF between the BLR radius and the optical luminosity measured between 1997 and 2021 ($r_{\rm max}=0.87$), 
and the blue dashed histogram is the cross-correlation centroid distribution (CCCD), the measured time lag of $4.6^{+1.7}_{-1.3}$~years is marked by the vertical dotted line in gray. 
}
\label{fig_stad}
\end{figure*}

\begin{figure}[ht!]
\centering
\includegraphics[angle=0,width=0.48\textwidth]{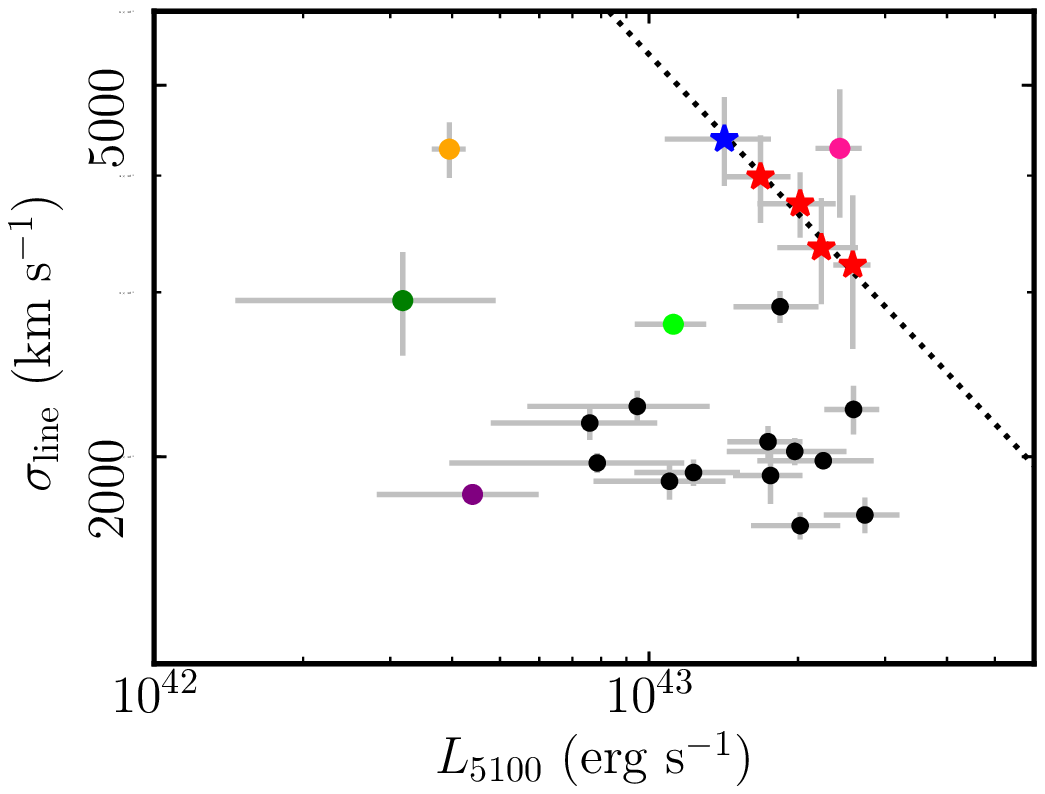} 
\caption{\footnotesize
Relation between the H$\beta$-based line dispersion ($\sigma_{\rm line}$) and optical luminosity of NGC~5548. 
The same symbols with Figure~\ref{fig_rl} and \ref{fig_stad} are used to distinguish the different campaigns. 
To guide the eye, a dotted line with a slope of $-0.5$ is plotted over the data from this work. 
}
\label{fig_widthbreth}
\end{figure}

\subsection{Stability of the BLR}\label{sec:stab}
The central ionizing source illuminates the BLR, which then emits broad emission lines through the photoionization process. 
Increased continuum radiation caused by a larger accretion rate will enhance the broad-line emissivity of BLR gases at larger distances, 
leading to the BLR radius (size) increasing and the line width decreasing with continuum luminosity, 
which is known as the normal breathing effect of BLRs (e.g., \citealt{Wang2020}). 
Currently, more than 100 changing-look$/$changing-state AGNs were found and 
identified (e.g., \citealt{Yang2018,Shu2018,Graham2020,Liu2021}). 
Those AGNs underwent extreme (large) variability 
and sometimes underwent spectral type transition between type~1 and type~2, 
are the good candidates for investigating the stability of BLR 
in different luminosity (accretion) states. 

For NGC~5548, we plot the optical luminosities at 5100~\AA\ ($L_{\rm 5100}$) and the BLR radius ($R_{\rm BLR}=c\tau_{\rm H\beta}$) 
as function of time in panels ({\it a} and {\it b}) of Figure~\ref{fig_stad}, respectively. 
The maximum difference of the optical luminosity is $\Delta {\rm log}~L_{\rm 5100}{\rm [erg~s^{-1}]} \thickapprox 0.93$ over the past thirty years, 
that is, the optical flux of NGC~5548 at high-luminosity state is $\sim$8.2~times larger than that at low-luminosity state. 
We find that the optical luminosity reduces to the lowest value around 2008, then gradually returns to the highest level around 2020, which also appeared around 1998 (see Figure~\ref{fig_stad}). 
The averaged BLR radius weighted by the emissivity also shows obvious variations in NGC~5548. 
The maximum difference of the BLR radius is $\Delta R_{\rm BLR}\thickapprox 22$~light-days. 
Inspecting the secular variations of the optical luminosity and BLR radius shown in Figure~\ref{fig_stad}, 
we can gather that the BLR radius follows the varying optical luminosity to some degree, 
however, it seems that the lowest BLR radius occurred around 2012, lagging the lowest state of the optical luminosity (around 2008). 

Using the same procedure for calculating the time lags of the broad emission lines in Section~\ref{sec:lag}, 
we find that, in NGC~5548, the change of the BLR radius lags the varying optical luminosity with a timescale of $\tau_{(\tau_{\rm H\beta,}L_{\rm 5100})}=3.5^{+2.0}_{-1.8}$~years. 
The results of cross-correlation analysis (in red solid curve) and corresponding CCCD 
(in blue solid histogram) are shown in panels~({\it ba}) of Figure~\ref{fig_stad}, 
where the measured time lag is marked by the vertical dotted line (in black). 
By adding our five-season RM results (displaying in the symbol of star), we find this time lag 
is longer than the value of 2.35~years reported by \cite{Lu2016}. 
Using the line width of broad H$\beta$,  we can estimate the BLR's dynamical timescale $t_{\rm BLR}=c\tau_{\rm H\beta}/V_{\rm FWHM, H\beta}$ (see also \citealt{Lu2016}). 
The averaged H$\beta$ lag ($\tau_{\rm H\beta}\sim12.3$~days) and averaged H$\beta$ line width ($V_{\rm FWHM, H\beta}\sim{\rm 6000~km~s^{-1}}$) 
over 23 RM campaigns yield a dynamical timescale of $\sim$2.1~years. 
As thus, $\tau_{(\tau_{\rm H\beta,}L_{\rm 5100})}$ is 
slightly longer than the BLR's dynamical timescale in NGC~5548. 

Interestingly, after undergoing extreme damping in the optical luminosity, 
the change of the BLR radius in NGC~5548 seemingly follows the optical luminosity in an even tighter fashion, 
which is more clearly seen in our five-season RM measurements 
(as if $\tau_{(\tau_{\rm H\beta,}L_{\rm 5100})} \sim 0$, 
see the symbol of star in Figure~\ref{fig_stad}). 
To make further investigation, a sub-set of the BLR radius and optical luminosity
after 1997 (marked by the vertical dashed lines in panel~{\it a} and {\it b} of Figure~\ref{fig_stad}) are selected to recalculate the time lag. 
The results of cross-correlation analysis (in red dashed curve) and corresponding CCCD (in blue dashed histogram) are over-plotted in panels~({\it ba}) of Figure~\ref{fig_stad}, 
where the measured time lag is marked by the vertical dotted line (in gray). 
This yields a relatively longer time lag of $\tau_{(\tau_{\rm H\beta,}L_{\rm 5100})}'=4.6^{+1.7}_{-1.3}$~years. 
Currently, the difference between $\tau_{(\tau_{\rm H\beta,}L_{\rm 5100})}'=4.6^{+1.7}_{-1.3}$~years 
and $\tau_{(\tau_{\rm H\beta,}L_{\rm 5100})}=3.5^{+2.0}_{-1.8}$~years is statistically insignificant, 
whether the BLR response timescale varies with the illumination history is unclear. 
Therefore, continuous RM campaigns with high-quality data are needed to understand the evolution of BLR radius and optical luminosity in NGC~5548. 

The relation between the line width of the broad H$\beta$ and the optical luminosity 
for NGC~5548 is investigated in Figure~\ref{fig_widthbreth}. This figure shows that the 
line dispersion of broad H$\beta$ from 23 RM campaigns does not globally correlate with the 
optical luminosity, which is not consistent with the \cite{Wang2020}'s finding from the 
SDSS-RM sample that the line width of broad H$\beta$ decreases with increasing optical luminosity 
(i.e, normal breathing). On the contrary, 
the line dispersion of the broad H$\beta$ from our five-season RM 
(showing in ``$\bigstar$") is inversely correlated with the optical luminosity.
In Figure~\ref{fig_widthbreth}, a dotted line with a slope of $-0.5$ is plotted over the data to guide the eye. 
The explanations for these discrepancies are unclear and more investigations are needed. 

In addition, extreme variability including AGN continuum and broad emission lines is 
the most significant observation feature of changing-look$/$changing-state AGN. 
The mainstream view is that a dramatic change in accretion rate triggers the changing-state 
process of AGN (e.g., \citealt{Sheng2017}). 
But a deeper question is what is the mechanism responsible for the dramatic change in accretion rate? 
Whether dramatic change occurs in the BLR during or before changing state. 
If the BLR supplies material to the accretion disc, 
the dramatic change in BLR structure and kinematics should occur ahead of 
the change in accretion rate for the changing-state case. 
For NGC~5548, although a changing-state process was not observed, 
the extreme variability that occurred in the past decades means that NGC~5548 is 
a potential candidate for changing-state AGN. 
Long-term and continuous RM observations for the BLR of NGC~5548 should be a crucial pathway to studying the intrinsic origin of the changing state in the future. 

\section{Black Hole Mass and Accretion Rate}\label{sec:mass}
The examination of the virial relation in Section~\ref{sec:vr} shows that, on a 30-years timescale, 
the BLR of NGC~5548 is dominated by the gravitational potential of the central SMBH. 
In this case, the SMBH mass can be well estimated by the virial equation 
\begin{equation}
M_{\rm BH} = f\frac{c\tau_{\rm H\beta} V^{2}}{G}, 
\label{eqn:mass}
\end{equation}
where $\tau_{\rm H\beta}$ is the H$\beta$ time lag, $c$ is the speed of light, $c\tau_{\rm H\beta}$ means the BLR radius, 
$G$ is the gravitational constant, 
the BLR velocity or line width $V$ is either FWHM or line dispersion ($\sigma_{\rm line}$) of 
the broad emission lines measured from the mean or rms spectrum, 
$f$ is a dimensionless virial factor, which incorporates the unknown geometry, kinematics, and inclination of the BLR. 
\citet{Ho2014} found that the uncertainty of virial factor $f$ can be reduced after considering the bulge types 
(including classical bulge and pseudobulge) of the host galaxy. 
For example, in case of line dispersion $\sigma_{\rm line}$ measured from rms spectrum, $f=6.3\pm1.5$ for classical bulges. 

In light of the factors that (1) 
the line-width measurement from the rms spectrum eliminates contamination of the constant 
components; (2) the line dispersion of $\sigma_{\rm line}$ produces less biased mass 
measurement than the FWHM (see \citealt{Peterson2011}); 
and (3) NGC~5548 hosts a classical bulge (\citealt{Ho2014}). 
We calculate the SBMH masses for our five-season RM campaigns and the previous 
RM campaigns in a uniform way, using the line dispersion $\sigma_{\rm line}$ of 
the broad H$\beta$ from the rms spectrum and $f=6.3\pm 1.5$. We also calculated the virial 
products defined as $c\tau_{\rm H\beta}\sigma_{\rm line}/G$ for all RM campaigns. 
The results are listed in Table~\ref{tab_sum}, 
where the last five rows list the results from our five-season RM campaigns. 
The SMBH masses derived from the previous 18 campaigns, ranging between 2.07 and 13.82 
in units of $10^7M_{\odot}$, are relatively diverse.  However, 
the SMBH masses derived from our five-season RM campaigns, ranging between 12.45 and 16.97 
in units of $10^7M_{\odot}$, are almost consistent with each other. 
Our five-season RM yields an averaged virial SMBH mass of $14.22\times10^7M_{\odot}$, 
with a small standard deviation of $1.89\times10^7M_{\odot}$. 
The stellar velocity dispersion of classical bulge in NGC~5548 is 
$\sigma_{*}=195\pm13~{\rm km~s^{-1}}$ (\citealt{Woo2010}). 
According to the $M_{\rm BH}-\sigma_*$ relation of \cite{Kormendy2013}, 
we obtained $M_{\rm BH}|_{\sigma_*}=(27.50\pm 8.80)\times 10^7M_{\odot}$ in \cite{Lu2016}. 
This SMBH mass is remarkably consistent with our measurements within uncertainties. 

The dimensionless accretion rate is defined as (\citealt{Du2015}) 
\begin{equation}
\dot{\mathscr{M}}=\frac{\dot{M}_{\rm BH}c^2}{L_{\rm Edd}}=20.1\left(\frac{\ell_{44}}{\cos i}\right)^{3/2}M_7^{-2}, 
\label{eqn7}
\end{equation}
where $\dot{M}_{\rm BH}$ is the mass accretion rate, 
$L_{\rm Edd}=1.5\times 10^{38}\left(M_{\rm BH}/M_{\odot}\right){\rm erg~s^{-1}}$ is the Eddington luminosity, 
$i$ is the inclination of the accretion disc, $\cos i=0.75$ is adopted, which represents a mean disc inclination for a type 1 AGN, 
$\ell_{44}=L_{5100}/10^{44}~{\rm erg~s^{-1}}$ is optical luminosity at 5100~\AA, 
$M_7=M_{\rm BH}/10^7M_{\odot}$ is the SMBH mass. 
With the optical luminosity and virial SMBH mass, the dimensionless accretion rates for each season are estimated and listed in Table~\ref{tab_sum}. 
The 23 RM campaigns yield an averaged accretion rate of $\langle \dot{\mathscr{M}} \rangle=0.034$ with a standard deviation of 0.034, 
indicating that NGC~5548 is overall in a sub-Eddington accretion state. 
During our five-season RM campaigns, 
NGC~5548 has the maximum accretion rate of 
$\dot{\mathscr{M}}=0.025$ in the season of 2020 (see Table~\ref{tab_sum}). 

\section{Summary} \label{sec:sum}
We began a long-term spectroscopic monitoring project for NGC~5548 in 2015, 
which aimed to investigate its BLR properties and accurately measure the virial SMBH mass. 
This project was undertaken with the Lijiang 2.4~m telescope. 
We had performed five-season RM observations between 2015 and 2021, 
with the median sampling interval ranging from 1.25 to 3 days. In this work, 
we conducted the basic measurements and carried out the velocity-resolved RM analysis. 
Spectral fitting and decomposition processes were employed during spectral analysis 
to improve the RM measurements. {\it For the seasons of 2015, 2018, 2019, 2020, and 2021}, 
we obtained the following main results. 
%\begin{itemize}
\begin{enumerate}
\item
We measured the light curves of the AGN continuum at 5100~\AA\ and the broad He~{\sc ii}, 
He~{\sc i}, H$\gamma$, and H$\beta$ lines. The time lags of these broad-line light curves 
with respect to the AGN continuum at 5100~\AA~were measured, ranging from 0 to 10~days, 
and their mean lags over the five seasons are 
$<\tau_{\rm He~II}>=0.69, <\tau_{\rm He~I}>=4.66, <\tau_{\rm H\gamma}>=4.60, <\tau_{\rm H\beta}>=8.43$ days. 
This demonstrates that the BLR in NGC~5548 obeys the radial ionization stratification. 
\item
We constructed the H$\gamma$ and H$\beta$ velocity-resolved lag profiles (VRLPs) for the seasons 
of 2015, 2018, 2019, and 2021, from which we found that both the H$\gamma$ and H$\beta$ VRLPs  
exist an `M'-shaped structure in the season of 2015 (also in 2014, see \citealt{Pei2017}), 
but this structure disappears after 2018. 
Infalling and Keplerian motion could dominate the kinematics of BLR in NGC~5548, 
and two kinds of motion might coexist, leading to the complex BLR kinematics. 
Alternatively, the accretion disc winds$/$outflows with different transparencies from the accretion disc partially shield the BLR, 
which may also give rise to complex signatures of VRLPs. 
We found that the H$\gamma$ and H$\beta$ VRLPs are varying with the seasons, implying an evolution in the kinematics of BLR. 
Continuous velocity-resolved RM experiments are needed to figure out the decisive explanation. 
\item
After eliminating other blended components and correcting for the instrumental broadening, 
we measured the FWHM and $\sigma_{\rm line}$ of broad H$\beta$, H$\gamma$, and He~{\sc i} 
from the mean and rms spectra for each season. 
Using the line dispersion of the H$\beta$ from the rms spectrum, the H$\beta$ time lag, and the 
dimensionless virial factor ($f_{\rm BLR}=6.3$) of the classical bulge, 
we calculated the virial SMBH mass of NGC~5548 from our five seasons of observations, 
which ranges from $12.45\times10^7M_{\odot}$ to $16.97\times10^7M_{\odot}$. 
The averaged virial SMBH mass is $M_{\bullet}/10^{7}M_{\odot}=14.22$, 
with a small standard deviation of $1.89$. 
%\end{itemize}
\end{enumerate}

By combining the previous 18 RM campaigns and our five-season campaign for NGC~5548, 
we obtained the following results and remarks. 
%\begin{itemize}
\begin{enumerate}
\item
We derived the $R_{\rm H\beta}-L_{\rm 5100}$ relation of NGC~5548 with a slope of 0.57. 
The mean values of $c\times\tau_{\rm H\beta}$ and $L_{\rm 5100}$ over the 23 campaigns are 
consistent with the canonical $R_{\rm H\beta}-L_{\rm 5100}$ relation. 
The measurements from our five-season RM overall lie $0.13$~dex below the canonical relation, 
but the resulting slope is consistent with the canonical relation (\citealt{Bentz2013}). 
Our results actually increase the weight below the canonical relation, 
making the slope of NGC~5548's individual $R_{\rm H\beta}-L_{\rm 5100}$ relation close to 0.5. 
\item
We obtained the virial relation of the BLR in NGC~5548 
and found that the whole BLR is bounded by the SMBH's gravitation well. 
The virial SMBH masses updated from the previous 18 RM campaigns range from 2.07 to 13.82 in units of $10^7M_{\odot}$, which are relatively diverse.  
As a comparison, the virial SMBH masses from our five-season RM, 
range from 12.45 to 16.97 in units of $10^7M_{\odot}$ with a much small scatter. 
\item
We found that the change of the BLR radius lags behind the change of the optical luminosity with a timescale of $3.5^{+2.0}_{-1.8}$~years. 
This timescale is relatively larger than the value (2.35~years) reported by \cite{Lu2016}, 
and longer than BLR's dynamical timescale of $\sim$2.1~years. 
However, after undergoing extreme damping in the optical luminosity, the BLR radius seemingly shows a 
normal breathing effect (i.e., the BLR radius increases with increasing optical luminosity), 
which is more clearly seen in our five-season RM measurements. 
To make further investigation, 
a sub-set of the BLR radius and optical luminosity measured after 1997 was selected to recalculate the time lag, 
which yields a longer timescale of $4.6^{+1.7}_{-1.3}$~years. 
Currently, the difference between both timescales is statistically insignificant, 
whether the BLR response timescale varies with the illumination history is unclear. 
In addition, we found that the line dispersion of the broad H$\beta$ from 23 RM campaigns is
not globally correlated with the optical luminosity, while line dispersion from our five-season RM is well inversely correlated with the optical luminosity. 
%\end{itemize}
\end{enumerate}

It is crucial to investigate the reasons for the above differences so as to better understand the structure and evolution of BLR. 
We will continue to monitor NGC~5548 for this purpose. 
The RM results of our long-term spectroscopic monitoring project for the other AGNs (e.g., Mrk~1018, Mrk~590, SDSS~J153636.22+044127.0) 
will be presented in future contributions. 

\section*{Acknowledgments}
We thank the referee for the useful report that improved the manuscript. 
This work is supported by the National Key R$\&$D Program of China with No. 2021YFA1600404, 
the National Natural Science Foundation of China (NSFC; 12073068, 11991051, 11873048, 11703077), 
and the science research grants from the China Manned Space Project with No. CMS-CSST-2021-A06. 
K.X.L. acknowledges financial support from the Yunnan Province Foundation (202001AT070069), 
the Youth Innovation Promotion Association of Chinese Academy of Sciences (2022058), 
%the Ten Thousand Talents Program of Yunnan for Top-notch Young Talents, 
the Top-notch Young Talents Program of Yunnan Province, 
and the Light of West China Program provided by Chinese Academy of Sciences (Y7XB016001). 

The long-term spectroscopic monitoring project of AGNs has gone through many years, we acknowledge the support of the staff of the Lijiang 2.4~m telescope. 
Funding for the telescope has been provided by Chinese Academy of Sciences and the People’s Government of Yunnan Province.

\end{document}